\documentclass[traditabstract]{aa}
\usepackage{graphicx}
\usepackage{txfonts}
\usepackage{natbib}
\begin{document}

\title{An E-ELT Case Study: Colour-Magnitude Diagrams of an 
Old Galaxy in the Virgo Cluster.}
 \titlerunning{Colour-Magnitude Diagrams in Virgo}

\author{A. Deep\inst{1,2}, G. Fiorentino\inst{1}, E. Tolstoy\inst{1},
E. Diolaiti\inst{3}, M. Bellazzini\inst{3}, P. Ciliegi\inst{3}, R.I. Davies\inst{4} \& J.-M. Conan\inst{5}
}
\authorrunning{Deep et al.}
\institute{
Kapteyn Astronomical Institute, University of Groningen, PO Box 800, 9700 AV Groningen, The Netherlands\\
\email{deep@strw.leidenuniv.nl, fiorentino@astro.rug.nl, etolstoy@astro.rug.nl}
\and
Leiden Observatory, Leiden University, NL-2300 RA Leiden, the Netherlands.
\and
INAF-Osservatorio Astronomico di Bologna, via Ranzani 1, I-40127 Bologna, Italy.
\and
Max-Planck-Institut f\"{u}r Extraterrestrische Physik, Postfach 1312, 85741 Garching, Germany 
\and
ONERA, 92322 Chatillon, France
}

\date{Received ; accepted }

\abstract{
One of the key science goals for a diffraction limited imager on an
Extremely Large Telescope (ELT) is the resolution of individual stars
down to faint limits in distant galaxies. The aim of this study is to
test the proposed capabilities of a multi-conjugate adaptive optics
(MCAO) assisted imager working at the diffraction limit, in IJHK$_s$
filters, on a 42m diameter ELT to carry out accurate stellar
photometry in crowded images in an Elliptical-like galaxy at the
distance of the Virgo cluster. As the basis for realistic simulations
we have used the phase A studies of the European-ELT project,
including the MICADO imager (Davies \& Genzel 2010) and the MAORY MCAO
module (Diolaiti 2010).  We convolved a complex resolved stellar
population with the telescope and instrument performance expectations
to create realistic images.  We then tested the ability of the
currently available photometric packages STARFINDER and DAOPHOT to
handle the simulated images.  Our results show that deep
Colour-Magnitude Diagrams (photometric error, $\pm$0.25 at I$\ge$27.2;
H$\ge$25. and K$_s\ge$24.6) of old stellar populations in 
galaxies, at the distance of Virgo, are feasible at a maximum surface
brightness, $\mu_V \sim$ 17 mag/arcsec$^2$ (down to M$_I > -4$
and M$_H \sim$ M$_K > -6$), and significantly deeper (photometric
error, $\pm$0.25 at I$\ge$29.3; H$\ge$26.6 and K$_s\ge$26.2) for $\mu_V
\sim$~21~mag/arcsec$^2$ (down to M$_I \ge -2$ and M$_H \sim$ M$_K
\ge -4.5$).  The photometric errors, and thus also the depth of the
photometry should be improved with photometry packages specifically
designed to adapt to an ELT MCAO Point Spread Function.  We also make
a simple comparison between these simulations and what can be expected
from a Single Conjugate Adaptive Optics feed to MICADO and also the
James Webb Space Telescope.
}


\maketitle

\section{Introduction}

It is hoped that in the not too distant future Extremely Large
Telescopes (ELTs) working at their diffraction limit will be available
\citep[e.g.,][]{Gil07, Szeto08, Johns08}.  A telescope with a mirror diameter of
$\sim$42m working at the diffraction limit will have a spatial
resolution of $\sim$10~mas in K$_s$ filter\footnote{In this paper
whenever the K filter is referred to it is always meant K$_s$.}  and
$\sim$5~mas in I filter.  These filters span the range of wavelengths
that are planned to be covered by (relatively) wide field ($> 1$arcmin
square) imagers making use of advanced multi-conjugate adaptive optics
(MCAO).  This will make possible a number of ground-breaking science
cases making use of the exceptional spatial resolution, and
sensitivity that is possible with such a large aperture telescope 
\citep[e.g.,][]{Najita02, Hook07, Silva07, KP09}.

One of the key science cases for an ELT is the imaging and
spectroscopy of individual stars in resolved stellar populations
\citep[e.g.,][]{Wyse02, Olsen03, Tolstoy10}.  It is one of the three
highlighted science cases for a European ELT presented by
\citet{Hook07}.  This science case is very broad and includes a range
of targets in the Local Group (including embedded star clusters within
the Milky Way) and also galaxies out to the Virgo cluster and beyond.
A primary goal is long held to be, to resolve individual stars in
Elliptical galaxies \citep[e.g.,][]{Olsen03} and to be able to
unambiguously interpret their luminosities and colours in terms of a
detailed star formation history and chemical evolution.  The nearest
predominantly old, classical large Elliptical, galaxies are to be
found in the Virgo Cluster.  There is a significantly closer example
of a peculiar Elliptical galaxy, Cen~A (at $\sim$3.4~Mpc, classified
as S0 by RC3) but it is certainly not representative of the class of
Elliptical galaxies.  NGC~3379 is an Elliptical galaxy at a distance
of 10.5~Mpc \citep{salaris98}  which is $\sim$0.7 magnitude ((m-M)$_0 \sim$ 30.3) less distant than
Virgo ((m-M)$_0 \sim$ 31), but it is a single system which may also
not be very typical \citep[e.g.][]{Capaccioli91}. The Virgo cluster
contains 2000 member galaxies \citep{Binggeli85}, 
with a range of morphology and luminosity, and also around 30 
``classical'' Elliptical galaxies.

A science case involving resolved stellar populations can be very
demanding because it typically requires excellent image quality (at
the diffraction limit) as well as optimum sensitivity.  Stellar
photometry requires accurate measurements over a large dynamical range
in two or more broad band filters covering as large a wavelength range
as possible.  The photometry of individual stars in a field of view
are then plotted in a Colour-Magnitude Diagram (CMD) the properties of
which depend upon the the star formation history and the chemical
evolution of the stellar system going back to the earliest times
\citep[e.g.,][and references therein]{Piotto05, Tolstoy09}.

Individual Red Giant Branch (RGB) stars have been photometered in very
long HST/ACS exposures ($\sim$~10~hrs per filter) of very low surface
brightness ($\mu_B > 27$) regions in diffuse dwarf spheroidal galaxies
in Virgo \citep[e.g.,][]{Caldwell06, Durrell07}.  The CMDs in these
studies contain very few stars, all within one magnitude of the tip of
the RGB, and with a large incompleteness and uncertainty due to
crowding. To hope to obtain deeper and more accurate photometry, and
also to be able to look at the large, bright classical Elliptical
galaxies in Virgo higher sensitivity and spatial resolution are
required.  As a bare minimum requirement this means detecting and
accurately photometering stars in crowded images at the tip of the RGB
(M$_I = -4$; M$_K \sim -6$) which means stars with I$\sim$27 and
K$\sim$25 at a signal-to-noise (S/N) $\sim 4$ at a surface brightness,
$\mu_V \sim 19$ mag/arcsec$^2$.  However, from the tip of the RGB
alone very little information about the star formation history of a
galaxy can be uniquely determined because of the well known
age-metallicity degeneracy.  It is important for a variety of reasons
to look deeper into the stellar population, and ideally reach the
Horizontal Branch (M$_I \sim$ M$_K \sim$ 0.)  which means
I$\sim$K$\sim$31, or even the oldest Main Sequence Turnoffs (M$_I =
+4$; M$_K = +3.$) which means I$\sim$35 and K$_s\sim$34. Detecting old
main sequence turnoffs is the most reliable way to determine an
accurate star formation history \citep[e.g.,][]{Gallart05}. However,
the Horizontal Branch, and to a lesser extent the shape of the red
giant branch also provide useful constraints on the ages and
metallicities of individual stars in a complex stellar population.

We can also hope to detect Infra-red (IR) luminous Asymptotic Giant
Branch and Carbon stars in Virgo Ellipticals, which will provide an
insight into the intermediate age stellar populations in these systems
\citep[e.g.,][]{Maraston06}. These stars are in many ways the ideal
targets of IR surveys, but they are not very representative of
the over all star formation history of a galaxy
\citep[e.g.,][]{Tolstoy10w}. They are only present for intermediate
age stellar populations, and even then the number is not clearly
determined by the star formation history alone.

Young stellar populations ($<$1~Gyr old), can also be very luminous and
will be easier to study than the old populations ($>$1~Gyr old) for
galaxies in the Virgo cluster and beyond.  Very young ($< 10$~Myr old)
massive stars are much brighter than their older siblings (e.g., M$_I
< -4$) but they are often buried deep in dusty molecular clouds under many
magnitudes of extinction and so clearly will benefit from IR observations
\citep[e.g.,][and references therein]{Tolstoy10}.

Here we aim to carry out simulations of resolved old
stellar populations in galaxies in the Virgo cluster based upon technical
information provided by Phase A E-ELT instrument projects.  Because of
the challenging demands on both sensitivity and spatial resolution,
detailed studies of resolved stellar populations require careful
simulations to understand if they are feasible.  We simulate a range
of surface brightness and determine how well standard photometry
packages are able to cope with image crowding and how this affects the
sensitivity and accuracy of the resulting CMDs.  We have chosen
not go into the detailed analysis of star formation histories coming
from the different simulations because our main interest is in the
photometeric accuracy that can be achieved. Naturally improved
photometric accuracy leads to more accurate star formation histories,
and this will be quantified in future work (Fiorentino et al., in prep).

Using the same simulation techniques we also compare our results to
those that may be expected from Single Conjugate Adaptive Optics
(SCAO) and the James Webb Space Telescope (JWST) in similar filters
for this science case.

\section{Creating \& Analysing Simulated Images}
\label{sec:create}

%Table 1
\begin{table}
\centering
%\begin{minipage}{80mm}
\caption{Telescope and instrument parameters for E-ELT, MICADO and MAORY.}
\begin{tabular}{@{}ll@{}}
\hline
Parameter  & Value \\
\hline
Collecting area		(m$^2$)	& 1275 \\
Telescope throughput 	&	0.74  \\
AO throughput	&	0.80 \\
Instrument throughput	 &	0.60\\
Total throughput 	& 	0.40 \\
(including detector QE) & \\
Read noise 	(e-)	& 5\\
Pixel Scale	(arcsec/pixel)	& 0.003\\
\hline
\label{table:tel_param}
\end{tabular}
%\end{minipage}
\end{table}

%Table 2
\begin{table}
\centering
%\begin{minipage}{100mm}
\caption{Filter characteristics in Vega magnitudes for MICADO.}
\begin{tabular}{@{}lllll@{}}
\hline
Filters  & I & J & H & K$_s$ \\
\hline
Filter center ($\mu$m) &	0.900	& 1.215	& 1.650	& 2.160 \\
Filter width	 ($\mu$m)	&0.24	&0.26	&0.29	&0.32\\
Zero magnitude 		&3.76$\cdot10^{10}$	&2.02$\cdot10^{10}$	&9.56$\cdot10^9$&	 4.66$\cdot10^9$\\
(ph/s/m$^2$/$\mu$m) &&&&\\
Background	 	&19.7 	&16.5 	&14.4 	&13.5 \\
(mag/arcsec$^2$)&&&&\\
Background	 	&0.6 	&5.8 	&20.9 	&25.7 \\
(e$^-$/s/pixel)&&&&\\
\hline
\label{table:filter_param}
\end{tabular}
%\end{minipage}
%\footnotesize{SR and EE values have been computed with a seeing=0.6~arsec.}
\end{table}

\begin{figure}
\centering
$\begin{array}{c}
\includegraphics[angle=0,width=9cm]{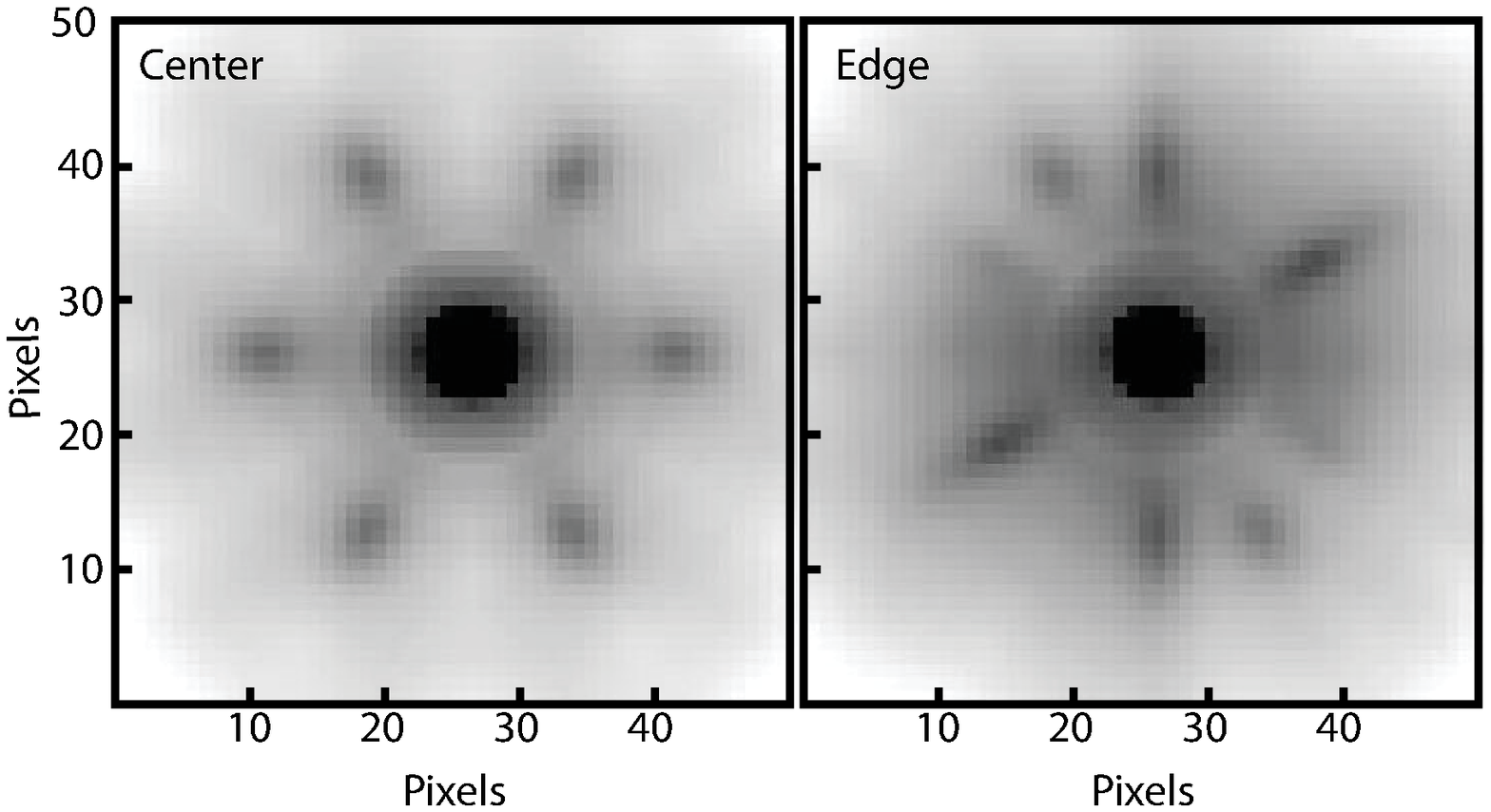} \\
\resizebox{\hsize}{!}{\includegraphics{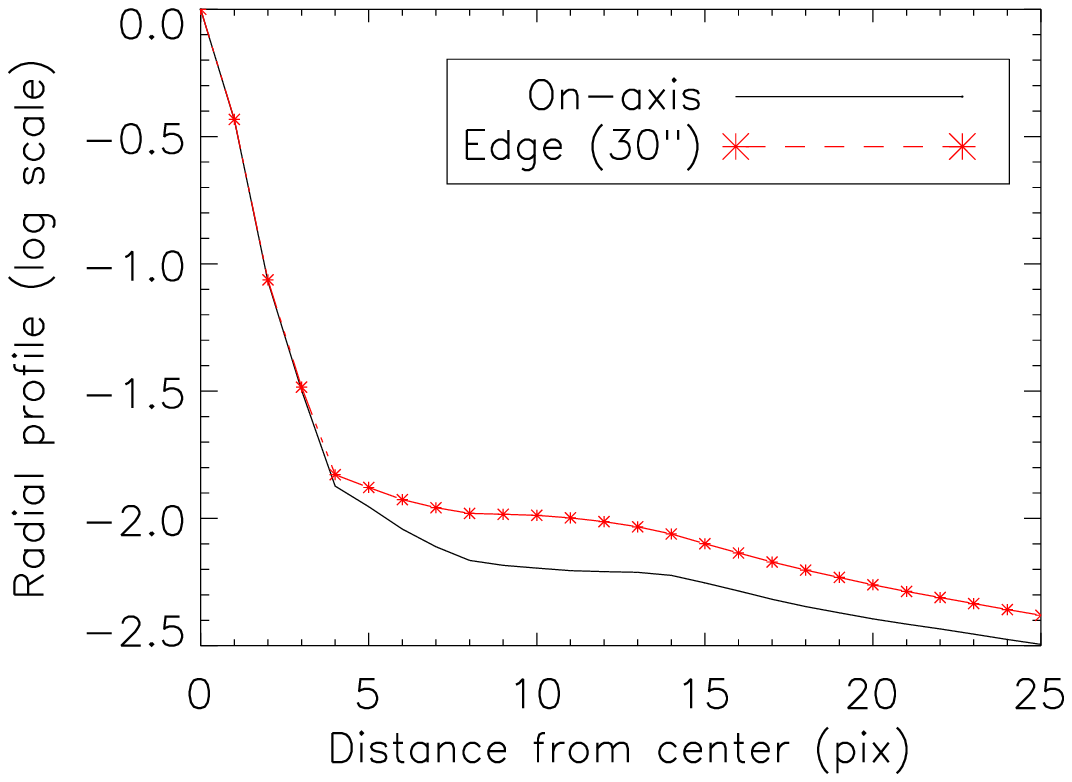}}
\end{array}$
 \caption{
We show the most challenging MCAO Point Spread Functions for the
optical I filter, provided by the MAORY consortium, created assuming a
seeing of 0.6~arcsec. One is at the centre, and one at a position
30~arcsec from the centre, at the edge of the field.  The images are
150~mas square, thus with a pixel scale of 3mas.  The averaged radial
profile compares the two PSFs over the same area as the images.
}
 \label{fig:PSFs}
\end{figure}

We create realistic simulations of images of crowded stellar fields in
Virgo cluster galaxies, using the technical specifications provided by
the European Extremely Large Telescope (E-ELT)
\footnote{http://www.eso.org/sci/facilities/eelt/} project based at
ESO \citep{Spy08} along with the E-ELT Phase A instrument study,
MICADO \citep{Davies10, Davies10a} \footnote{
http://www.mpe.mpg.de/ir/instruments/\\micado/micado.php?lang=de} and
the AO facility
MAORY\footnote{http://www.bo.astro.it/\~{}maory/Maory/}
\citep{Diolaiti10, Foppiani10}, see Table~\ref{table:tel_param},
Table~\ref{table:filter_param} and Fig.~\ref{fig:PSFs}.  Our goal is
to determine if it is possible to obtain useful CMDs of resolved
stellar populations in distant galaxies, specifically Elliptical-type
galaxies, at the distance of the Virgo cluster (17~Mpc).  This
question cannot be answered using simple estimates of sensitivity and
resolution because of the complex shape of the PSF (see
Fig.~\ref{fig:PSFs}) and the extremes of image crowding expected.
Simulated images are also necessary to assess the difficulties in
carrying out accurate photometry with standard packages, given the
complex PSF shape. Therefore simulations have been performed to create
realistic images, varying the input assumptions to understand the most
important effects on the photometric accuracy and depth.

%Figure 1

\subsection{The Instrument}
\label{subsec:inst}

The MICADO imager is proposed to cover a 53 arcsec square field
of view at the diffraction limit of the 42m E-ELT in I, J, H and K$_s$
broad band filters, with a pixel scale of 3~milli-arcsec/pixel
(3~mas/pixel).  It is foreseen to be fed by the MAORY MCAO module,
which is based upon 6 Laser Guide Stars in a circle 2~arcmin in
diameter around the field centre, for the high order sensing. It also
uses a number of Natural Guide Stars to measure the low order
modes. This facility easily includes the MICADO field of view.  The
key advantage of MCAO is to increase the size of the field corrected
and also the uniformity of the correction over the field. MICADO is
also able to use a SCAO (single-conjugate adaptive optics) module,
making use of natural guide stars, in the early phase of its operation
\citep{Clenet10}, although this reduces the field of view to 45~arcsec
diameter (in K$_s$), with a rapidly decreasing performance towards the
edge of this field, and also with decreasing wavelength.

Photometry of resolved stellar populations has to date been
predominantly carried out in optical filters (e.g., B, V \& I or
similar). This is certainly the preferred wavelength range to carry
out accurate and deep photometry of individual stars on the Main
Sequence and in cases of low interstellar extinction.  This is partly
because this corresponds to the peak of the luminosity distribution of
the main sequence stars, and also partly because the sky background is
relatively faint and stable.  Excellent image quality has been
obtained for deep optical exposures of resolved stellar populations at
stable ground-based sites with active image correction (e.g., Paranal,
Las Campanas \& Mauna Kea) and most notably from space, with the
Hubble Space Telescope.  MICADO/MAORY on the E-ELT is foreseen to be
most efficient at near-IR wavelengths, as this is where the AO
performance is best. The only optical images which we can hope to
obtain, with even a minimum acceptable AO performance, are through the
I filter.  This is also the shortest wavelength for which a MAORY PSF
has been provided, and it has been provided with a number of caveats,
the most important being that it is at the edge of what is possible to
achieve with this post-focal adaptive optics mode/configuration
combined with this telescope.  The other broad band filters available
are in the near-IR, namely, J, H and K$_s$, and they are projected to
have a much better AO performance. These filters are commonly used to
study regions of heavy interstellar extinction (e.g., the Galactic
Bulge, and its globular clusters), and they will remain useful to
significantly extend this work.

We have carried out simulations for all the broad band filters over
the full wavelength range MICADO/MAORY proposes to operate.  Our aim
is to understand the effects of a peculiar AO PSF on the photometric
results at all wavelengths, combined with the varying background
level, diffraction limit, Strehl and crowding in all the different
filters.

\begin{figure}
%\centering
\resizebox{\hsize}{!}{\includegraphics{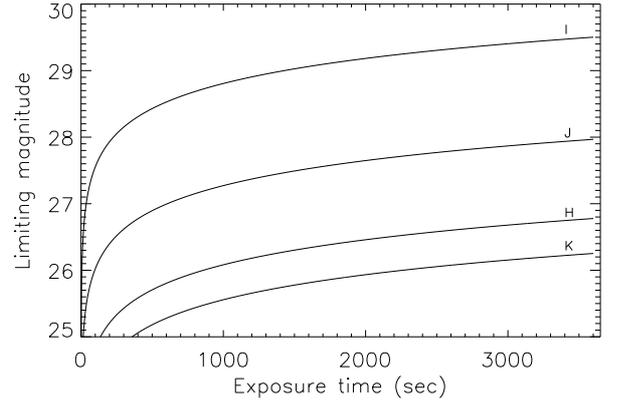}}
 \caption{
Limiting magnitudes for I, J, H and K$_s$ filters (Vega magnitudes) as a
function of time for S/N=4 calculated over a 50~mas aperture, assuming
a seeing of 0.6~arcsec.
}
 \label{fig:lim_mag}
\end{figure}

To predict the sensitivity of the telescope and instrument combination
we used the MAORY PSFs to estimate the theoretical limiting magnitude
that can be achieved with MICADO/MAORY as a function of time for S/N$=
4$ using a 50~mas diameter aperture (see Fig.~\ref{fig:lim_mag}).
This value of S/N corresponds to an error of $\pm$0.25 in magnitude
(using $\sigma_{error} = -2.5\log(1+1/(S/N))$.  These theoretical
values will be compared with the results of our simulations. From
Fig.~\ref{fig:lim_mag} it can be seen that the limiting (Vega)
magnitudes are I$\sim$29.5, J$\sim$28.0, H$\sim$26.8 and K$_s\sim$26.2
for an exposure time of 1 hour. This is the exposure time adopted for
all simulations presented here.  We have also assumed cold nights with
almost no thermal background to 2.32$\mu$m.  These conditions,
although optimistic, are realistic and should be available for around
two months every year. This assumption will have almost no effect on
the I sensitivity, but a large effect on K$_s$, and a decreasing effect on
H and J.

%Figure 2

\subsection{The Point Spread Function}
\label{subsec:psf}

The importance of a well defined Point Spread Function for accurate
stellar photometry is well documented \citep[e.g.][]{Stetson87,
Schechter93}.  To be able to carry out reliable photometry, which can
be accurately calibrated, down to faint magnitude limits it is
necessary to be able to correctly model the brighter stars and remove
them from the image to see the fainter stars below the ``wings'' of
the brighter stars.  Accurate crowded field photometry of individual
stars on MCAO corrected images is challenging mostly because of the
complex PSF (Point Spread Function), with a sharp central core
surrounded by an extended diffuse halo (see Fig.~\ref{fig:PSFs}),
which has a strong effect on the crowding properties of stellar
images.  Both these components of the PSF will vary in time and also
across the field of view due to anisoplanatism.  The most accurate way
to assess the capabilities and the impact of an usual PSF shape over
the wide field of view with an MCAO imaging system is to simulate the
expected images as realistically as possible and then to analyse them
using standard techniques.

Thus the most crucial aspect of making realistic simulations is to
have an accurate estimate of the form and variation of the PSF.  The
MAORY consortium estimated the MCAO PSF modelling atmospheric and
instrumental effects \citep{Diolaiti10}.  Two PSFs in the I filter are
shown in Fig.~\ref{fig:PSFs}, one at the field centre and one at the
edge of the field.  The I PSF is the most technically challenging, for
the MCAO system, and potentially the most important for stellar
population studies of the kind considered here.
Fig.~\ref{fig:lim_mag} suggests that it will go significantly deeper
in than J, H or K$_s$.  These PSFs, which were last updated in March 2010,
and include all major sources of error, like the cone effect due to
LGS and the Natural Guide Star Wave Front Sensor errors.  

\begin{figure}
\resizebox{\hsize}{!}{\includegraphics{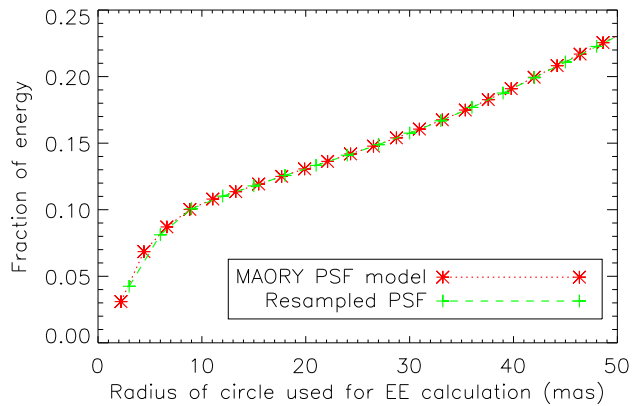}}
 \caption{
For the I PSF, the Encircled Energy (EE) is plotted as a function of
distance from the center for the original MAORY PSF (red star symbols)
and the PSF resampled to the MICADO pixel scale (green crosses).
}
 \label{fig:resampling}
\end{figure}

The PSFs have been calculated by sampling the diffraction limit at the
Nyquist limit (approximately 2 pixels per FWHM). This means that all
the PSFs provided have the same absolute size (512 $\times$ 512
pixels), and thus the pixel size differs with wavelength, and all have
to be resampled to match the 3~mas pixel scale of MICADO.  It can be
seen that the radial average of the PSF in the lower panel of
Fig.~\ref{fig:PSFs} is smoothed by the MICADO sampling.  To obtain
correct results, the resampled PSF should have a similar energy
distribution to the original, especially near the centre.  This is
important because the variation in encircled energy (EE) is very steep
in the centre making it difficult to extrapolate and any error has a
major effect as the photometry is typically strongly weighted by the
the central pixels.  After resampling, we check that the PSF energy
distribution matches that of original (see Fig.~\ref{fig:resampling}),
and as can be seen the original and the resampled PSF have almost
exactly the same EE.

%Figure 3

Two main parameters which are commonly used to evaluate AO image
quality, are the Strehl ratio (SR) defined as the ratio between the
peak intensity of a measured point source and the ideal diffraction
limited case; and the Encircled Energy (EE), defined as the fraction
of the energy enclosed in a circular aperture of a given diameter
about a point source.  The mean values of SR and EE for the MAORY PSFs
used in this work are summarised in Table~\ref{table:ee-sr}.  The EE
values have been estimated for a 50~mas aperture.  These EE values are
used for the estimates of the absolute sensitivity, shown in
Fig.~\ref{fig:lim_mag}.  The mean values of SR and EE shown in
Table~\ref{table:ee-sr} decrease when moving from the centre to the
edge ($\sim$25~arcsec from the centre) of the field, by an amount that
increases more for shorter wavelengths. Table~\ref{table:ee-sr} shows
that MAORY achieves a high degree of uniformity and is stable over the
field of view, especially for longer wavelengths (e.g., K$_s$).

%Table 3
\begin{table}
\centering
%\begin{minipage}{100mm}
\caption{The expected mean value for Encircled Energy (EE), 
for a diameter of 50~mas, 
and Strehl Ratio (SR) from MAORY, computed with a seeing of 0.6~arcsec.}
\begin{tabular}{@{}lllll@{}}
\hline
Filters  & I & J & H & K$_s$ \\
\hline
SR(center)	 	&0.065	&0.22	&0.44	&0.62 \\
SR(edge)	 	&0.05	&0.20	&0.41	&0.60 \\
EE(center)	 	&0.14	&0.30	&0.48	&0.62 \\
EE(edge)	 	&0.11 	&0.26	&0.44	&0.59 \\
EE(seeing=0.8)	 	&0.08 	&0.22	&0.41	&0.56 \\
FWHM (pix) & 1.76 & 2.03 & 2.33 & 2.60 \\
\hline
\label{table:ee-sr}
\end{tabular}
%\end{minipage}
\end{table}

%Figure 4
 \begin{figure}
\resizebox{\hsize}{!}{\includegraphics{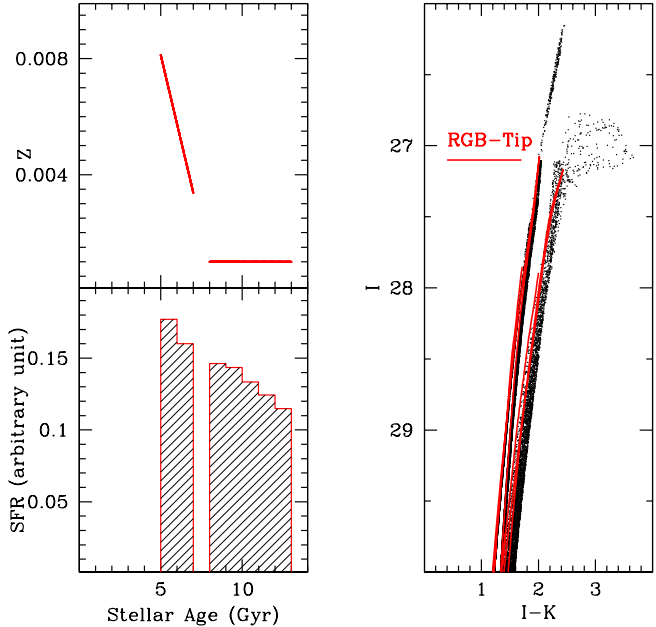}}
 \caption{
Input properties of the stellar population for a predominantly old
galaxy. On the left, the star formation rate with time
is shown in the lower panel and the corresponding chemical evolution
is shown in the upper panel. The resulting CMD of the galaxy is shown
on the right, created using IAC-star \citep{Aparicio04}.  To highlight
the mean properties of the stellar populations, we have added
isochrones (red solid lines) to the CMD for an old stellar population
(10 Gyr) with Z=0.001 and an intermediate one (6 Gyr) with Z=0.004,
from \citet{Piet04}.
}
 \label{fig:science}
\end{figure}

\subsection{The Stellar Population}
\label{subsec:stellar_pop}

The defining characteristic of Elliptical galaxies is a highly
concentrated stellar population. Thus the main restriction to obtain
deep accurate CMDs of the central regions of distant Elliptical
galaxies is stellar crowding, or the number of stars expected per
resolution element. To properly test the capabilities of an E-ELT
imaging instrument to photometer stars accurately in these crowded
fields it is important to create a realistic model for the stellar
population that could be ``observed'' in these simulations. This
means that the numbers, colours and magnitudes of stars are
distributed as might be expected in a galaxy-like complex
stellar population.  The input population should include not only
those stars that are expected to be detected, but also the much larger
number of undetected faint stars that create unresolved background
fluctuations that can significantly affect the detection limit and the
accuracy of the observations.

The primary aim of these simulations is to understand the most
important effects that limit the accuracy of stellar photometry with
MICADO/MAORY. Thus we concentrate on a single stellar population, a
fixed CMD, and we simply vary the total number of stars
distributed in the images depending upon the surface brightness.  We
have chosen a stellar population that matches what might be expected
for a predominantly old galaxy at the distance of the Virgo cluster
(17~Mpc), see Fig.~\ref{fig:science}.

We have given the artificial stellar population, and hence the CMD we
are going to use for all our simulations, distinct features (see the
right hand side, Fig.~\ref{fig:science}) which allow the eye to pick
out how crowding and photometric errors effects the results. This
is not meant to mimic an expected stellar population, but to allow an
easy assessment of the effects of increasing errors on the photometric
sensitivity and fidelity.  These two distinct episodes of star
formation include one ancient and long lasting (extending from 13~Gyr
ago to 8~Gyr ago) plus a slightly younger (from 7~Gyr ago to 5~Gyr
ago) and more metal rich population (see left panels in
Fig.~\ref{fig:science}). These populations were created using IAC-STAR
\citep{Aparicio04} using Teramo stellar evolution libraries
\citep{Piet04}; bolometric corrections libraries \citep{Castelli01}
and we assumed a Salpeter Initial Mass function.  The stellar
populations in Fig.~\ref{fig:science} come directly from the stellar
models, including the E-AGB population above the tip of the RGB.  We
created a ``complete'' stellar population down to 0.5~M$_\odot$ for
our chosen star formation history.  For both components the
models also include a bright E-AGB population above the RGB, which is
predicted to be ubiquitous in Elliptical galaxies. These AGB stars are
easy to confuse with the RGB if the photometry is not sufficiently
deep and accurate, and the distance to the galaxy is not well known.
Even though this population is very luminous it can be very sensitive
to crowding effects due the substantial underlying stellar population
\citep[e.g.,][]{Stephens03}.

The population we have chosen is arguably more metal poor than might
be expected from a giant Elliptical galaxy, but this  is the more
challenging scientific case, as the RGB is more narrow and blue than
for a solar metallicity population. This case can also be
considered to provide a limit on our ability to detect and study a
metal poor population in a Virgo Elliptical galaxy. Changing this
population for a more metal rich example would not change any of the
results on the photometric accuracy presented here, as the relative
number of stars will remain approximately the same in a similar CMD at
higher metallicity, and only the colour distribution on the RGB will
change.

%Figure 5
\begin{figure} \resizebox{\hsize}{!}{\includegraphics{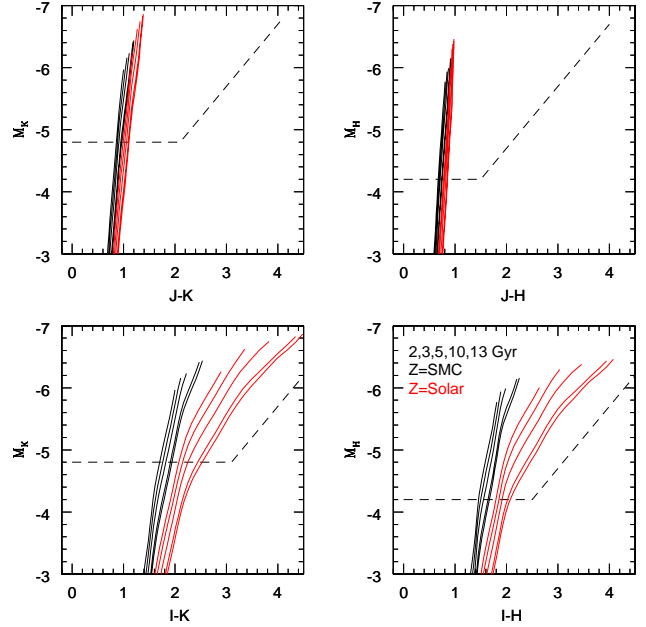}}
 \caption{
Here we show a selection of 5 isochrones \citep[from][]{Yi01} for
SMC-like metallicity (in black) and solar metallicity (in red) for the
ages 2, 3, 5, 10 \& 13 Gyr old. Each panel shows a different filter
combination that is possible with MICADO/MAORY. The dashed line comes
from the sensitivity estimates for one hour exposures.
}
 \label{iso}
\end{figure}

The differences expected on the RGB for different metallicity stellar
populations in a CMD are highlighted in Fig.~\ref{iso}, where it
can be seen that the results of our simulations will not be
significantly effected by a more metal rich population. The RGB stars
will all always lie within the sensitivity limits, although the very
red isochrones (the old, metal rich stars) are closer to the
sensitivity limits in I, and thus their photometry may be less
accurate at optical wavelengths.  Thus Fig.~\ref{iso} suggests that
the IR filters are likely to be more useful to study the most metal
rich stellar populations, and the optical-IR for metal poor stellar
populations. It is important for a realistic picture of the
entire evolution of any large galaxy that both metal rich and metal
poor stellar populations are accurately surveyed and their properties
quantified.

\subsection{Creating the Images}
\label{subsec:simulation}

The stellar population described in Section~\ref{subsec:stellar_pop}
is combined with the instrument and telescope parameters for different
filters described in Sections~\ref{subsec:inst} and \ref{subsec:psf}
to create realistic images in IJHK$_s$ filters.  The images have been
created using the IRAF (Image Reduction and Analysis Facility) task
{\it{mkobject}}. This task takes a list of stars, with magnitudes and
colours taken from our model stellar population, and using the MAORY
PSF to define their structure, randomly places the required number of
stars, to create the desired surface brightness, over the images.  The
parameter {\it{zeropoint}} determines the absolute number of photons
in an image for a star of a given magnitude in a one hour
integration. This parameter takes into account the area of telescope,
the through-put of the instrument and the distance of the star.  The
appropriate Poisson photon noise and read-noise for the detectors are
also added (see Table~\ref{table:tel_param}).

\begin{figure} \resizebox{\hsize}{!}{\includegraphics{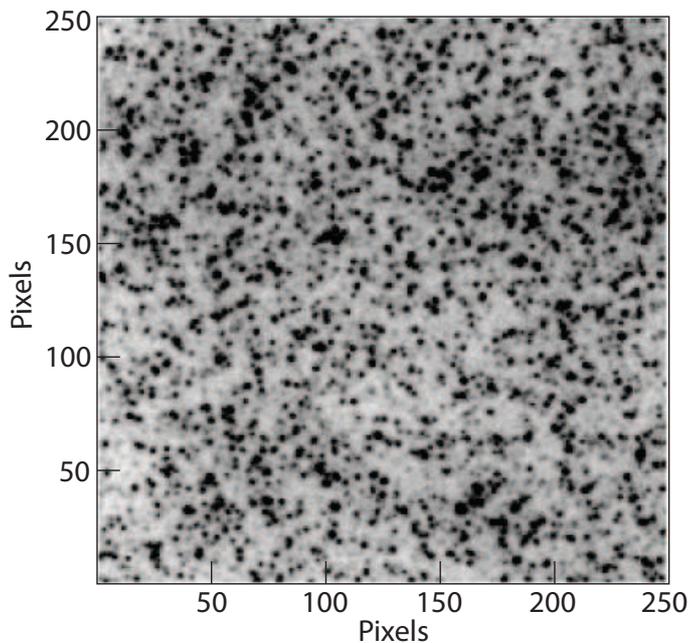}}
 \caption{
A simulated MICADO/MAORY image of a stellar field in an old galaxy at
the distance of Virgo (17~Mpc) in the I filter covering a field of
view of 0.75~arcsec square, with a pixel scale of 3~mas.  The surface
brightness of the galaxy at this position is $\mu_V \sim$19.0
mag/arcsec$^2$.  The image assumes an exposure time of 1~hour.
}
 \label{fig:sim_image}
\end{figure}

We always create 0.75~arcsec square images, which corresponds to 250
$\times$ 250 pixels.  This field is a very small fraction of the full
MICADO field, but it allows us not to worry about a varying PSF within
a single simulation and limits the time required to carry out the
large number of simulations and the corresponding analysis. Each image
is created with a constant PSF. This PSF will vary depending where
this image fragment is presumed to lie in the full MICADO field.  We
actually distribute the stellar population over a still larger area
(500 $\times$ 500 pixels), to ensure that the effect of the wings of
the PSF of stars outside the primary field will not be
under-estimated, as this would create an artificial ``edge-effect'' in
the final images.  Fig.~\ref{fig:sim_image} shows an example of one of
our simulated MICADO/MAORY images.

%Figure 6

Our simulations are the most realistic we could make given the
available information. There are however some simplifications we have
made that may lead to some differences with what will be delivered by
a real instrument.  The PSF shape in our images will vary depending
where it is placed on a pixel, and different programmes have different
ways to define this. We used two different interpolation methods, one
to place the PSFs and another to find and photometer them. This
results in an error of $\pm$0.03mag on the measured flux in our
photometry.  This will not be an effect in the real observations, so
this is an error that will artificially inflate the error budget of
our simulations, and we have to take this into account when we are
analysing the true accuracy of our photometry.

We also always assume a single one hour exposure time for each image.
This is certainly unrealistic, since in practice very much shorter
exposures will be taken and co-added.  Co-adding many images may result
in a slightly broader and smoother PSF, which may serve to minimise
some of the differences between real and simulated PSFs.  However, it
can be hoped that future instrument pipelines will be better able to
handle the PSF and how it varies. This is likely to develop from an
improved ability to model the atmospheric effects, and from PSF
reconstruction techniques.  Specifically there is likely to be a much
better defined PSF with a well mapped out time dependency.  We are
confident that our simulations realistically show what can be expected
of MICADO/MAORY, from the currently available technical
specifications.

\subsection{Photometry}
\label{sec:photometry}

To detect and measure the magnitudes of the stars in our simulated
images, we primarily used Starfinder \citep{Diolaiti00}, a photometry
package that was originally, like all others, developed to perform PSF
photometry on images with a constant PSF \citep{Diolaiti00}. It was
subsequently successfully adapted for the strongly distorted stars in
crowded fields in SCAO images \citep{Origlia08}. Initial attempts have
also been made to deal accurately with MCAO images from MAD
(Fiorentino et al., 2011, submitted). In this case the PSF does not
vary as strongly as for SCAO, but the variation is much more difficult
to model, in the case of MAD because of a strongly non-uniform
variation over the field of view coming from compromises made in the
natural guide star orientation and brightness.

Starfinder works by creating a 2D image of the PSF using bright
isolated stars from the observed field, without any analytic
approximations.  This PSF is then used as a template, which can be
scaled and translated by sub-pixel offsets, to detect stars by
matching their profiles against the template. The PSF theoretical PSF
can also be given directly to Starfinder if it is well known, to
perform photometry. We used the approach which is currently more
realistic of determining the PSF on image, as it would be for ``real''
observations.  The extracted PSF was compared with that used to make
the images and is found to be an excellent match to the original input
PSF.

%Figure 7
\begin{figure*}
\centering
 \includegraphics[width=17cm]{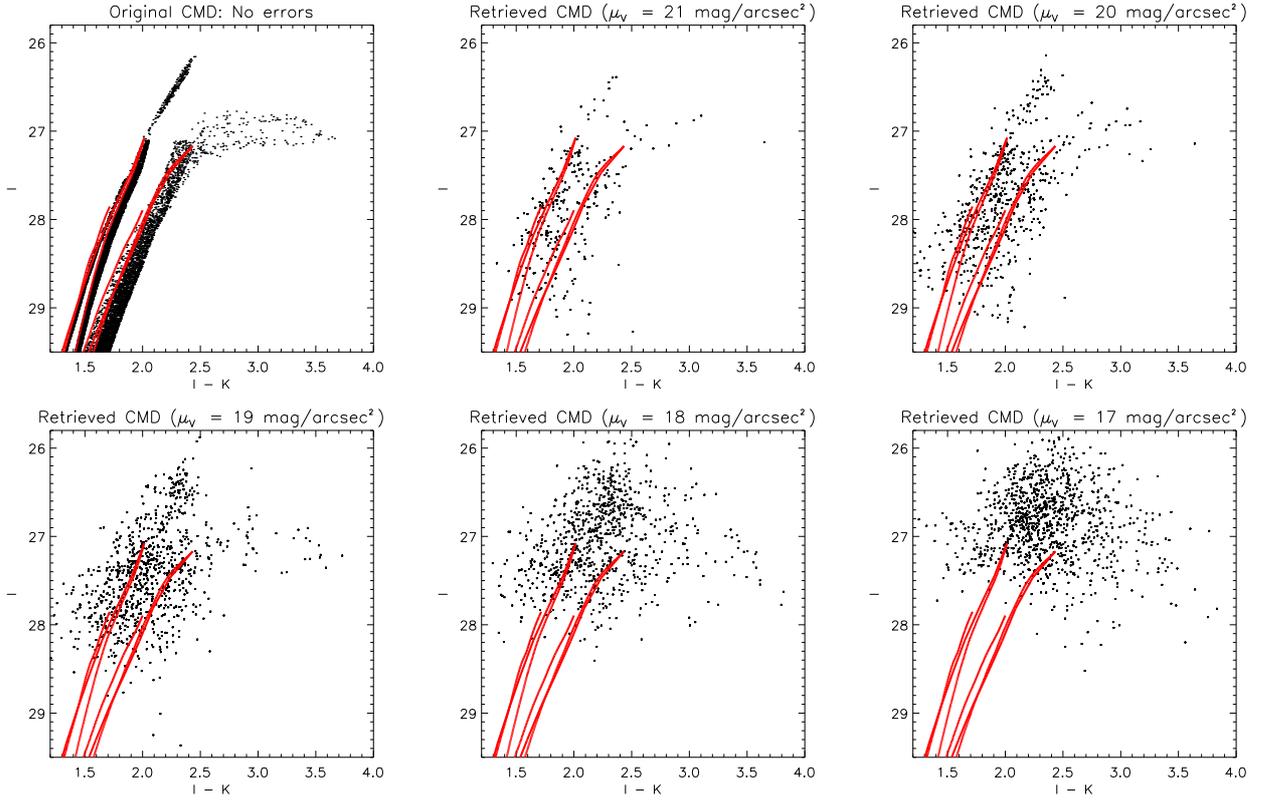}
 \caption{
The (I, I-K) CMDs obtained from Starfinder photometry of simulated
MICADO/MAORY images for 5 different surface brightness ($\mu_V$)
levels.  The top left hand corner is the original input CMD, without
any errors.  Included on each panel are the isochrones which represent
the mean of the properties of each distinct stellar population, namely
Z=0.001; 10~Gyr old and Z=0.004; 6~Gyr old.  The highest surface
brightness ($\mu_V = 17$~mag/arcsec$^2$) is equivalent to a distance
of $\sim$5~arcsec from the centre of a typical Elliptical galaxy in
Virgo (e.g., NGC~4472), whereas the lowest ($\mu_V =
21$~mag/arcsec$^2$) is at a distance of $\sim$75~arcsec from the
centre.
}
 \label{fig:cmd_elt_k}
\end{figure*}

A detection threshold of 3$\sigma$ above local background is taken as
the limit to which objects can be detected. First the brightest stars
are detected and removed from the images, and then the images are
searched again, iteratively, to find as many faint sources as possible
in a crowded stellar field.  This iteration is necessary to resolve
crowded groups, down to separations comparable to the limits allowed
by the width of the PSF.

\subsection{Surface Brightness \& Crowding}
\label{subsec:crowding}

The accuracy and the faint limit of the photometry from MICADO/MAORY
images towards the high surface brightness 
centres of Elliptical-like galaxies in Virgo will
be strongly affected by the extreme crowding of unresolved 
stars.  This is partly because of the large low surface brightness
halo surrounding the core of each PSF, but also simply due to the
extremely dense stellar population and the fact that we can only resolve
the brightest stars.  The higher the surface brightness
the more densely packed will be the stellar population and the higher
will be the unresolved background.  This effect is well known, and has
been the subject of much study over the years, starting in radio
astronomy \citep[e.g.][]{Scheuer57, Condon74} and more recently also
in optical and IR studies of crowded stellar populations
\citep[e.g.][]{Gallart96, Renzini98, Olsen03, Stephens03}. 

The extended halo of the MCAO PSFs which are hard to model accurately
in current photometry packages makes crowding effects particularly
acute.  A poorly subtracted background, which limits the detection of
fainter stars lying below the PSF wings of brighter stars is the
result.  Thus, how the surface brightness of an image relates to the
magnitude and crowding limits for detecting individual stars is what
needs to be quantified in our simulations.  The complex PSF, as well
as the significant underlying stellar population means that the
photometric accuracy is not a simple relation between the number of
stars and the number of pixels, although this does, of course, provide
a hard limit. It is important to verify the difference between the
theoretical magnitude limits and what can be achieved with real images
and real photometry.

In all cases most stars in our simulated images lie below the
detection threshold.  We found that stars more than 2 mag below the
detection threshold only contribute to the background level as a
uniform flux, and so they were added as such.  The rest of the stars
were added individually, as the fluctuations due to the marginally
detected stars, just below the detection threshold. These have a
significant effect on the photometric accuracy of the resolved stellar
population, especially those stars just above the detection threshold,
as they form highly variable background fluctuations.

To study crowding effects the density of the stellar population is
varied from 20~000 to 500~000 stars (going 2~magnitudes below the
detection threshold in each filter) per 0.75 arcsec square image.
In fact, for this particular stellar population, at this
distance, the background of unresolved stars rises up faster, with
increasing surface brightness, than the crowding limit of detected
stars. The numbers of stars put into an image of course correspond
to a surface brightness ($\mu_V$), and we have chosen to test five
different values: $\mu_V$ = 21, 20, 19, 18 and 17 mag/arcsec$^2$.  The
limits were chosen such that there remained statistically meaningful
numbers of stars detected with reasonably accuracy in each simulated
image. In the case of the lowest surface brightness ($\mu_V = 21$)
this relates directly to the low stellar density, but for the bright
limit ($\mu_V = 17$) the high flux in the unresolved background limits
the number of stars that can be accurately photometered.

\section{The Results: Colour-Magnitude Diagrams}
\label{sec:cmd}

From the simulated images in the four broad-band filters (I, J, H and
K$_s$) which have been photometered with Starfinder at five different
surface brightness levels we obtain numerous CMDs.  For example the
results for (I,~I-K), which covers the longest colour baseline, are
shown in Fig.~\ref{fig:cmd_elt_k}, for surface brightness values
between 21~mag/arcsec$^2$ and 17~mag/arcsec$^2$.  The effect of
increasing crowding on photometric depth and accuracy can be clearly
seen. We have chosen not to carry out a detailed star formation
history analysis on each simulation here, but it can clearly be seen
that the two distinct populations on the RGB merge more and more into
an indistinct blob, and the faint stars disappear as the surface
brightness and hence the crowding increases.  The two distinct input
stellar populations can barely be distinguished in the highest surface
brightness CMDs. This is mostly due to the increasing background due
to ever larger numbers of unresolved stars, but also partly due to the
increased crowding of the detected stars.  Severe crowding can make it
difficult to accurately measure the magnitudes of even the brightest
stars.

%Figure 8
\begin{figure} \resizebox{\hsize}{!}{\includegraphics{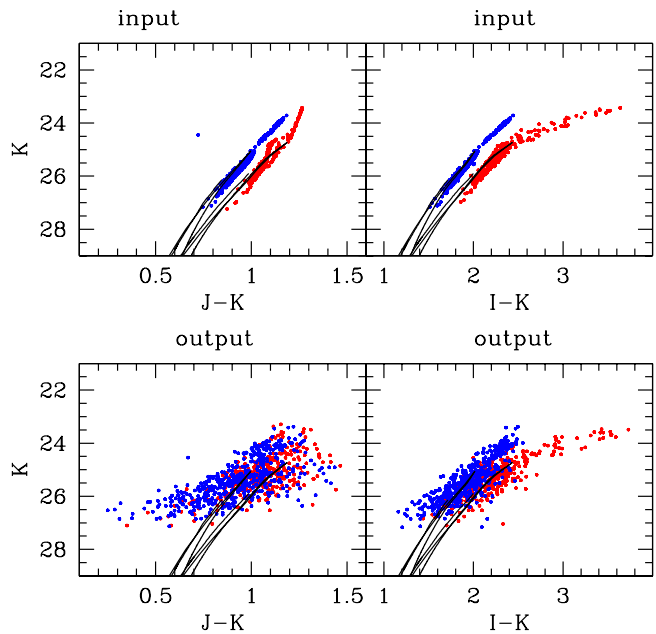}}
 \caption{
Here we show the stellar population that is the input to the simulated
images (upper panels) and the output photometry measured (lower
panels), in (K, J-K) and (K, I-K), assuming a surface brightness,
$\mu_V = 19$~mag/arcsec$^2$, for the stellar population defined in
Fig.~\ref{fig:science}. The two distinct star formation episodes are
colour coded, so that the impact of the different filter choices can
be better judged.
}
 \label{lim-cmds}
\end{figure}

To make the most efficient use of telescope time we should determine
which is the best combination of the available filters to obtain the
most sensitive and accurate CMDs which can be interpreted with the
least ambiguity. It would seem that the combination (I, I-K) leads to
the most detailed CMD, as can be seen from a set of isochrones (e.g.,
Fig.~\ref{iso}), and comparing them with the sensitivity limits
(Fig.~\ref{fig:lim_mag}) that the best combination is likely to be I
and K$_s$ filters. This is because the colour is more spread out when
the I filter is included, and the colour range of the isochrones is
broader.  However, this has to be tested with our simulations, which
will give us the most reliable indication of the E-ELT performances on
real data.  The filter combination with the best AO correction is (K,
J-K), and there are examples in the literature where the optical-IR
CMDs contain the same details as IR CMDs alone \citep[e.g.,][for Omega
Cen]{Sollima04} despite the compression of the colour scale.  Thus to
test the difference between J-K and I-K CMDs we have colour coded the
two populations in our simulations and compared the two CMDs at the
same surface brightness (19 mag/arcsec$^2$) in Fig.~\ref{lim-cmds}.
It can be seen that the output (``observed'') CMD is better defined in
(K, I-K), specifically the upper parts of the two RGB branches and the
AGB stars can be more clearly distinguished.  This is mostly due to
the small colour difference between these two populations, which can
barely be separated in the input CMD, with no measurement errors.

It is perhaps surprising that I-K is a better combination than J-K, as
the AO correction in the I band is very poor. However for the science
case we have chosen the poor AO correction in I is compensated by the
low sky background (see Table~\ref{table:filter_param}).  This will
not be true for very red stars, such as Carbon stars and metal rich
AGB stars which are intrinsically brighter in the IR. It will also not
be true where there is severe reddening, and in this IR photometry
alone will be able to penetrate the dust and produce accurate and deep
CMDs.

In summary, Fig.~\ref{lim-cmds} shows that although (K, I-K) is
preferable; with carefully modelling (K, J-K) can still provide
valuable information about the colour and the spread of the RGB and
AGB populations.

In our simulations we always assume that reddening is very low, and we
concentrate our effort on the properties of the I and K$_s$ filters as
these provide the deepest and most accurate CMDs of RGB stars for the
type of galaxy and stellar population we have chosen.

%Figure 9
\begin{figure*}
\centering
 \includegraphics[width=17cm]{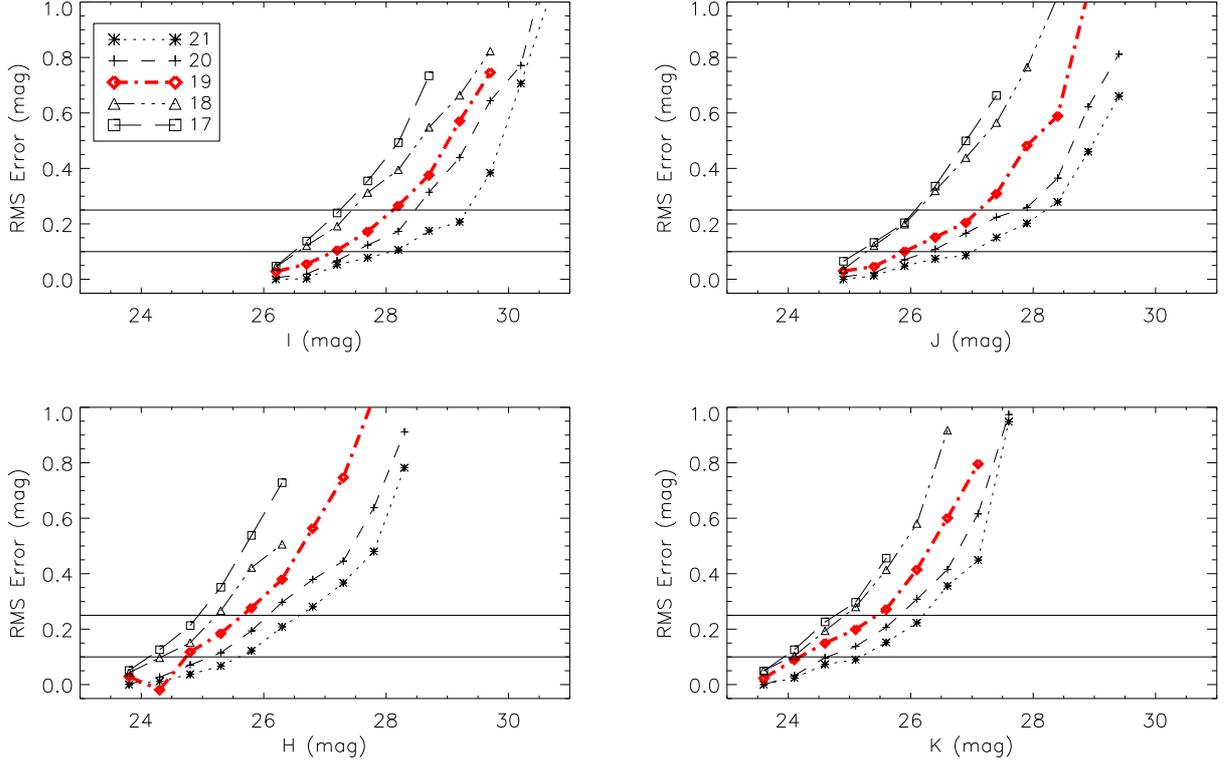}
 \caption{
Photometric errors, defined as the difference between the input and
the retrieved magnitudes, for simulations of MICADO/MAORY images in I,
J, H \& K$_s$ filters.  The results are given for 5 different surface
brightnesses ($\mu_V$), corresponding to the CMDs in
Fig.~\ref{fig:cmd_elt_k}.  Highlighted (in red) is the case for a
surface brightness, $\mu_V$ = 19 mag/arcsec$^2$, in each filter. This
corresponds to a distance of around 25 arcsec from the center of a
typical giant Elliptical galaxy, like NGC 4472.
}
\label{fig:errors_elt}
\end{figure*}

The two most important effects that dominate our ability to carry out
an accurate scientific analysis of deep images of resolved stellar
populations are the photometric errors and the completeness of the
photometry, and of course they are intricately dependent on each
other.  Errors determine how accurately we can distinguish between
different stellar evolution models, and hence age and metallicity of
stars in a complex stellar population. The completeness, or fraction
of the stars of a given magnitude that are detected, is important
because stellar evolution models make predictions about the relative
numbers of stars of different magnitudes. This means it is important
to know the fraction of a population that is detected to be able to 
distinguish between evolutionary effects measurement uncertainties.

%Figure 10
\begin{figure} \resizebox{\hsize}{!}{\includegraphics{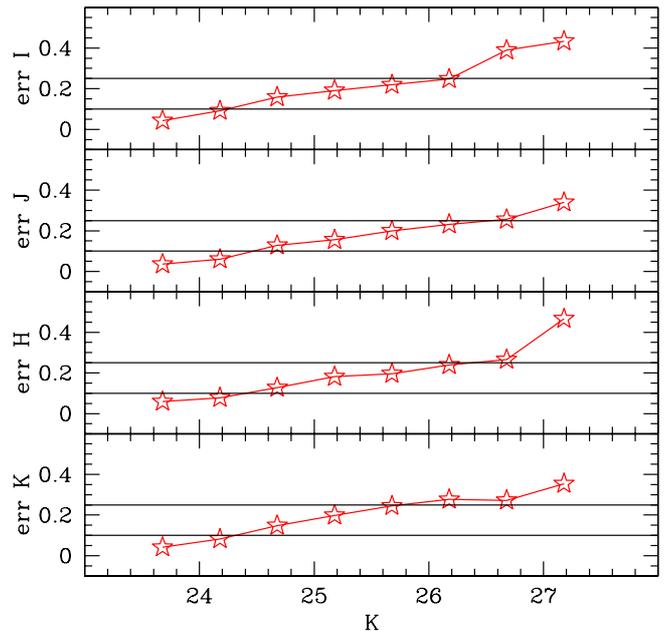}}
 \caption{
Here we show for the I, J, H, K$_s$ errors from our photometry, as a
function of K$_s$ magnitude. This shows the limiting factor on the
photometric accuracy of the CMD as a function of K$_s$ magnitude.
}
 \label{errors-k}
\end{figure}

\subsection{Photometric errors}
\label{sec:phot-errors}

Our simulations allow the ultimate test of how well the photometry is
carried out because we know a priori the properties of the stellar
population being photometered. This means that we can accurately
quantify how well and how complete the photometry can be done in a
range of different conditions. Input and output catalogues are matched
and stars are considered to be detected if they are found within 0.5
pixel of their original position, and $<$1.0 mag of the input
magnitude.  We computed the RMS error for each magnitude over a bin of
0.5 mag and these values are plotted for different $\mu_V$ in
Fig.~\ref{fig:errors_elt}.  These RMS values are prone to statistical
errors due to the finite numbers of stars in each bin. Magnitude bins
that have fewer stars will have a larger RMS error.  To mitigate this
effect we put 50 test stars in every magnitude bin and based our error
estimates solely on these test stars.  It is reassuring to see that
for the least crowded images the limiting magnitude and error in
Fig.~\ref{fig:errors_elt} corresponds well to the predictions in
Fig.~\ref{fig:lim_mag}.

As expected, the photometric errors become steadily larger towards
fainter magnitudes in all filters and for all surface brightness
values in Fig.~\ref{fig:errors_elt}.  The effect of increasing surface
brightness is to make the photometric error at a given magnitude
larger. It can also be seen that even at the brightest magnitudes the
photometric error is always larger than would be typical for HST or
high quality (non-AO) ground based studies.  But these are of course
inflated by the errors due to the different PSF interpolation methods
(see section~\ref{subsec:simulation}), The limitations of the
photometry packages which are not yet optimised for the kind of
extremely extended and irregular MCAO PSF (see Fig.~\ref{fig:PSFs}) is
also an important factor. This is clearly an area of development,
independent of instrument hardware, that needs to be undertaken before
optimum use can be made of resolved stellar photometry from an
instrument like MICADO/MAORY.

From Fig.~\ref{fig:errors_elt}, it can be seen that 
at a surface brightness, $\mu_V$ = 19 mag/arcsec$^2$, which implies a
distance of around 25 arcsec from the center of a typical giant
Elliptical galaxy, like NGC~4472, for stars of magnitude I$\sim$28.2,
J$\sim$27.2, H$\sim$25.7 and K$_s\sim$25.4 we can achieve a photometric
accuracy of $\pm$0.25 mag.  If we want to probe closer to the centre
of a giant Elliptical galaxy, say within 5~arcsec ($\mu_V$ = 17
mag/arcsec$^2$), we can do that for stars with magnitudes I$\sim$27.2,
J$\sim$26.2, H$\sim$25.0 and K$_s\sim$24.6 or brighter, with the same
error.  From Fig.~\ref{fig:errors_elt}, as was predicted in
Fig.~\ref{fig:lim_mag}, the I magnitude clearly has a much fainter
limiting magnitude than J, H or K$_s$, because of the lower sky background
in I.  This would suggest that the limiting factor in obtaining a deep
CMD with MICADO/MAORY is the IR magnitude.  However, the fact that the
same photometric error is obtained at I=28.2 and K$_s$=25.4 does not
necessarily mean that I is always better than K$_s$. This strongly depends
on the colour of the star being observed. This difference is actually
optimal for a star with I-K$\le$2.8. If the star is any redder than
this (I-K $> 2.8$) then the sensitivity of the I filter begins to be
the limiting factor.  In Fig.~\ref{errors-k}, we show the contribution
to the errors on the colours of the RGB in our CMDs as a function of
K.  It can be seen that at the limit of the K$_s$ sensitivity, the I error
is actually dominating the uncertainty in the I-K colour.  This
is because I has to be significantly more
accurate than K$_s$ to match accuracy of photometry for the same RGB star.

\subsection{Completeness}

\begin{figure*}
\centering
 \includegraphics[width=17cm]{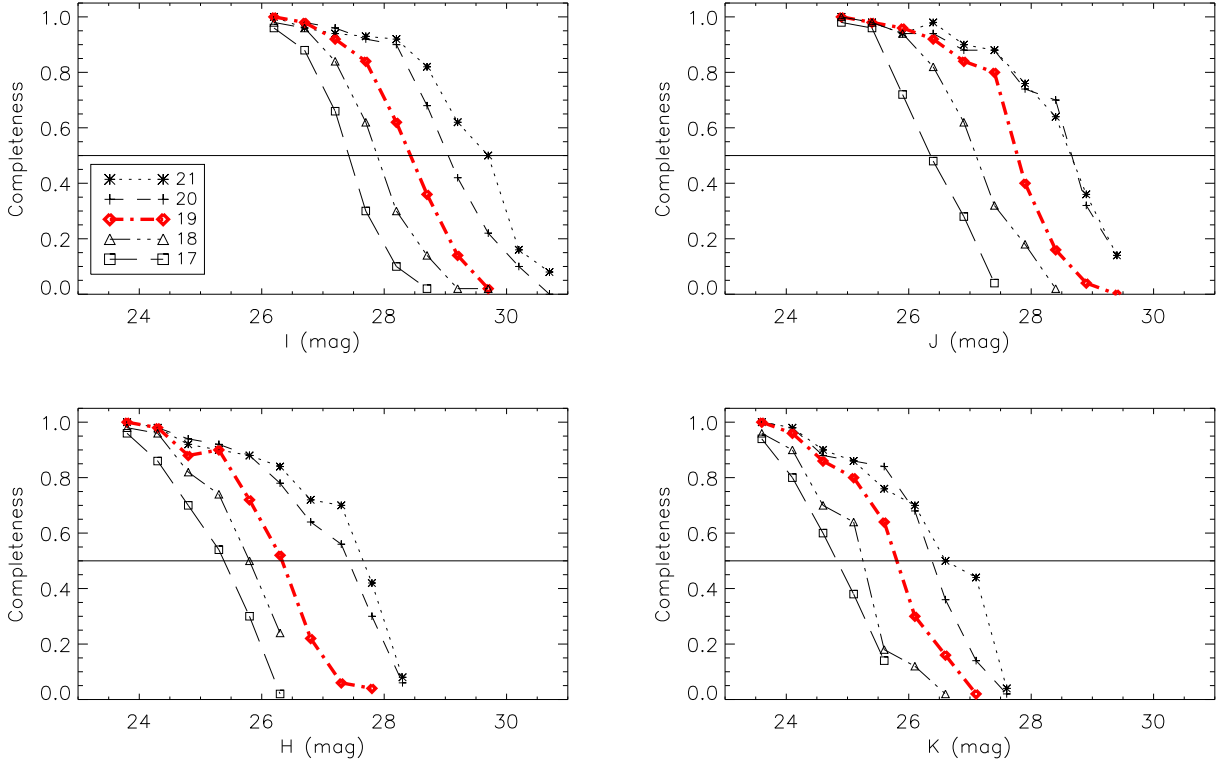}
 \caption{
Completeness fractions, which are defined as the fraction of input
stars retrieved in the output catalogues at given magnitudes, for
MICADO/MAORY images in I, J, H and K$_s$ filters. The results are
given for five different surface brightness ($\mu_V$) Highlighted (in
red) is the case for a surface brightness, $\mu_V$ = 19
mag/arcsec$^2$, in each filter, which implies a distance of around 25
arcsec from the center of a typical giant Elliptical galaxy, like
NGC~4472.
}
 \label{fig:completeness}
\end{figure*}

A fraction of the stars in an image are always, for a variety of
reasons, not detected. This is called the incompleteness
fraction. This effect can also be quantified in our simulated images,
where, as for ``real'' observations, this has to be carefully
quantified before the CMDs can be accurately interpreted.  The most
challenging regime is for stars which are only slightly above the
threshold of detection, that is $\sim 3\sigma$ above the background.
To estimate the completeness, or the typical fraction of stars which
are detected at a given magnitude, in our simulations we can simply
compare our input and output catalogues for all filters.  This is
broadly the same procedure used for any imaging study, except that in
this case {\it all} the stars are by definition added to the images,
so we obtain a much more direct {\it measure} of completeness.  The
results are shown in Fig.~\ref{fig:completeness}. In each case we have
assumed that a star is ``found'' if it is detected within $\pm$0.5
pixel from its original position and with a conservative magnitude
difference of $<$1~mag.  From Fig.~\ref{fig:completeness} it can be
seen that bright stars are almost always retrieved even in high
surface brightness images but the level of completeness declines for
fainter stars, at a rate that accelerates with increasing surface
brightness.

We also note the number of ``false'' detections, that is stars
incorrectly identified in the images.  We find that there are more
incorrectly identified stars in the I images, than in the K$_s$ images.
This is most likely because I has a very poor AO correction compared
to K$_s$ (see Table~\ref{table:ee-sr}).  Nevertheless the I filter is
still more sensitive to detect and photometer RGB stars than J, H or K$_s$
filters; and despite the worse AO correction, the completeness is
comparable in I and J, and both are better than H and K$_s$.  However, as
for the photometric errors the appropriate comparison is the I and K$_s$
completeness with an offset in magnitude related to the colour on the
RGB at that magnitude. 

\subsection{Natural Seeing}

The Natural Seeing assumed for the simulations of the MAORY PSF of
course have an effect on the shape of the PSF.  The MAORY consortium
has produced PSFs for natural seeing of both 0.6 and 0.8~arcsec.  As
expected, even though it is a small effect, I images are more affected
by seeing conditions than K$_s$. For 0.6~arcsec seeing, a photometric
accuracy of 0.25 is achievable for I$\sim$28.1 star. With 0.8~arcsec
seeing, the same accuracy will be achieved only for stars with
I$\sim$27.7 mag. There is no effect at bright magnitudes, and
especially in K$_s$ the effect at all magnitudes is small.

\subsection{The Effect of a Different Stellar Population}

If we look at a younger galaxy, still forming stars, 
or a galaxy closer by (or further
away) the main effect is going to be, for the same surface brightness,
a different level of background fluctuations.  This is because the
characteristics of the underlying stellar population, that which is
below the detection level, will change.  For example, nearby galaxies
will be more fully resolved into stars than distant galaxies for the
same observing time.  This means that the effects of an unresolved
stellar background will become less, and sensitivity and crowding
limits determined for a distant galaxy will be an over estimate for a
nearby galaxy.  If we look at nearby galaxies it becomes increasingly
important to take wide field images to sample the variation in a
galaxy (e.g., a spiral galaxy) or to get sufficient stars to properly
populate a CMD (e.g., nearby dwarf galaxies) at all magnitudes.

%Figure 11

Large numbers of bright very young stars will also have an effect on the
photometric sensitivity and crowding, because relatively high
numbers of bright stars make the contamination of the fainter
population with PSF wings more severe.  This
effect is stronger in I filter than in the IR filters, which is due to
the poor AO correction in I.

%Figure 12
\begin{figure}
\resizebox{\hsize}{!}{\includegraphics{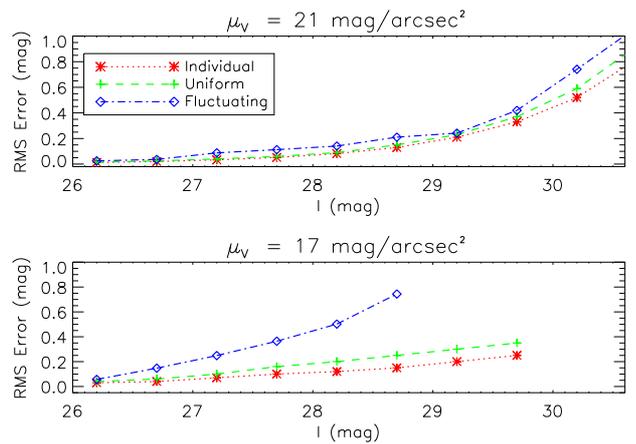}}
 \caption{
The effect of the type of background fluctuations on photometric
accuracy, for two different surface brightness, $\mu_V = 21$ (upper
panel) and $\mu_V = 17$ (lower panel). In each panel three cases are
considered for the same stars detected: individually; on a uniform
background those detected on a fluctuating background. They are
labeled, respectively, theoretical (red); uniform brightness (green);
fluctuations (blue).}
\label{fig:no_crowding}
\end{figure}

In Fig.~\ref{fig:no_crowding} we compare photometric accuracies in I
achieved at the same surface brightness for a nearby galaxy (uniform
background) and more distant (fluctuating background) stellar
populations.  As can be seen the presence of different background
fluctuations can have a significant impact on photometric accuracy
at high surface brightness.

\subsection{Comparison with DAOPHOT/ALLSTAR photometry}
\label{subsec: daophot}

We made a comparison between our Starfinder results at $\mu_V \sim 20$
mag/arcsec$^2$ with the commonly used DAOPHOT/ALLSTAR photometry
package \citep{Stetson87}. It uses a different approach to define the
PSF, and also includes the possibility for the PSF to vary over the
field of view.  DAOPHOT/ALLSTAR models the PSF using the sum of an
analytic bi-variate symmetrical function and an empirical look-up
table providing corrections to this function. This PSF is defined by
comparing the observed brightness values and the average profile of
numerous stars over the image.  This hybrid PSF offers flexibility in
modelling complex PSFs, even when AO is used.  It has also been tested on
MCAO images from MAD (e.g., Fiorentino et al. 2011, submitted) which
typically have a strongly varying PSF over the field of view.

%Figure 13
\begin{figure}
\resizebox{\hsize}{!}{\includegraphics{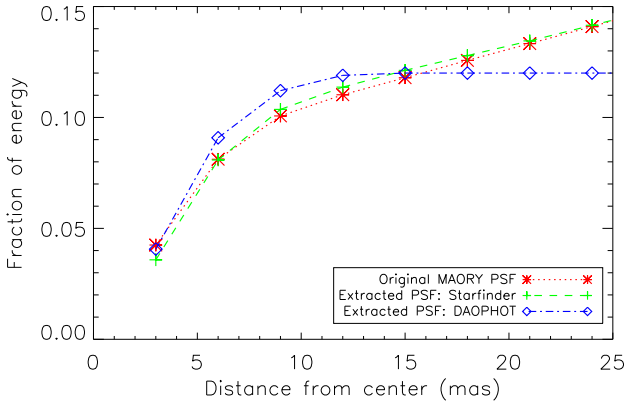}}
 \caption{ 
A comparison between the encircled energy as a function of distance
from centre of the field in the I filter, for the original MAORY PSF
(red); the PSF extracted by Starfinder (green); and that extracted by
DAOPHOT (blue).
}
\label{fig1:daophot_comparison}
\end{figure}

For the DAOPHOT PSF determination we selected $\sim$100 isolated stars
in a low surface brightness image to estimate an analytical PSF. This
allows a careful mapping of the PSF variations across a field.  We
then left DAOPHOT free to choose the best fitting form for the PSF,
allowing for a quadratic positional change. A comparison of the EE
distribution for the PSFs modeled by DAOPHOT with both the theoretical
MAORY PSF and that extracted from the image by Starfinder is shown in
Fig.~\ref{fig1:daophot_comparison}.  We can clearly see the difference
in the PSFs determined by DAOPHOT and Starfinder, where the Starfinder
PSF is closer to the shape of the input PSF.  This is because
Starfinder is extracting the PSF directly from the image, whereas
DAOPHOT makes a model.  However, when we integrate the flux, in both
cases the end results are very similar.

%Figure 14
\begin{figure*}
\resizebox{\hsize}{!}{\includegraphics{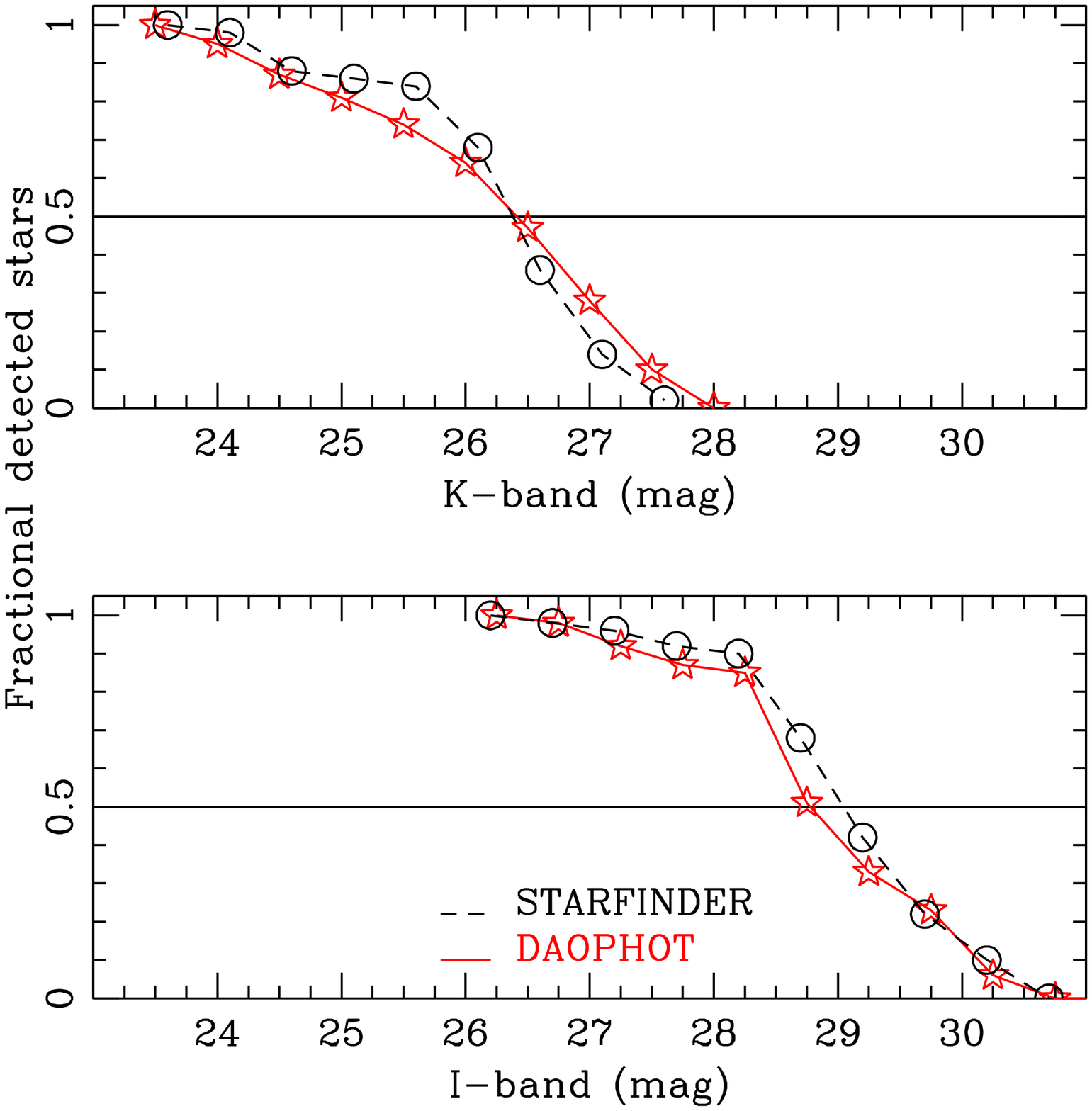}
\includegraphics{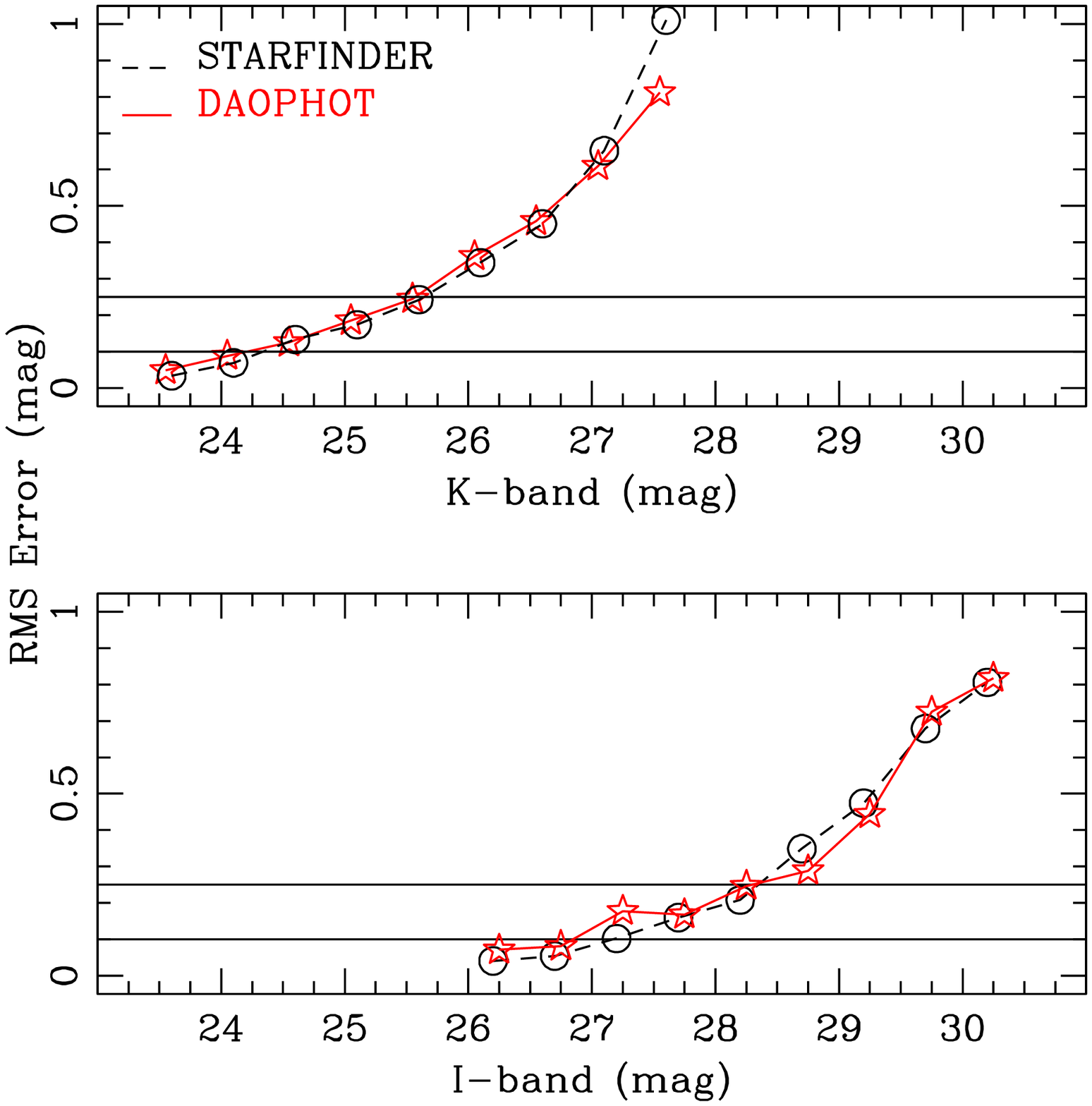}}
 \caption{
Comparison between the I and K$_s$-band completeness and photometric
errors from Starfinder (black) and DAOPHOT/ALLSTAR (red) for our
MICADO/MAORY images in I and K$_s$ at $\mu_V =$20 mag/arcsec$^2$.
}
\label{fig:daophot_comparison}
\end{figure*}

We used ALLSTAR to perform photometry, and
Fig.~\ref{fig:daophot_comparison} shows the comparison between the
photometry resulting from the two different packages. We compared the
completeness and the photometric accuracy.
Fig.~\ref{fig:daophot_comparison} shows that despite the differing
accuracy in modelling the PSF, the photometry obtained from DAOPHOT is
similar in terms of both accuracy and completeness, to the results
obtained from Starfinder for both I and K$_s$ photometry.  However,
given that Starfinder more accurately reproduces the PSF shape, it is
likely to be more accurate in more crowded stellar fields.

There is clearly room for improvement in existing photometric packages
to properly account for complex MCAO PSF, that should lead to more
accurate photometry especially at the fainter magnitudes.  This is
unlikely to affect the absolute detection limits but it should make
the errors smaller for fainter magnitudes which will make a
significant improvement in our ability to interpret CMDs.

%Figure 15
\begin{figure}
\resizebox{\hsize}{!}{\includegraphics{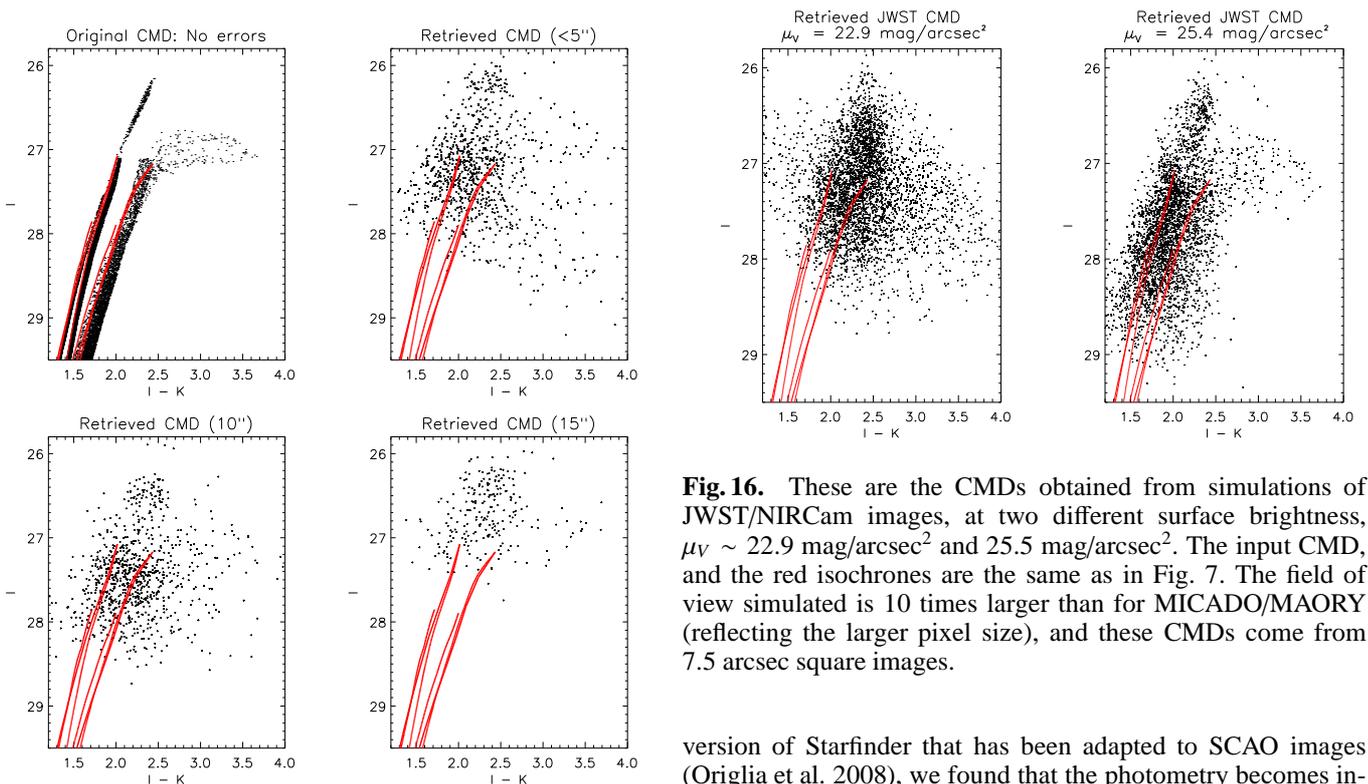}}
 \caption{ 
Here we show the effect of anisoplanatism for 0.75 arcsec square 
MICADO/SCAO images, with a
surface brightness, $\mu_V$ = 19 mag/arcsec$^2$ for a range of
distances off-axis from the single guide star. The input CMD, and the
isochrones are the same as in Fig.~\ref{fig:cmd_elt_k}.
}
\label{fig:scao}
\end{figure}

\subsection{Single Conjugate Adaptive Optics (SCAO) }

Another option for correcting images for a fluctuating atmosphere
which is being considered for (early) use with MICADO is
single-conjugate adaptive optics (SCAO). This uses only one guide star
(often a natural guide star) to measure the wavefront phase. In this
case, the corrected field of view is limited by the position of
science target with respect to the guide star, and to some degree also
by the properties of the guide star.  The anisoplanatism is very
significant in this case, especially for I filter observations.

We have performed simulations for the same stellar population used
through out this paper (see Fig.~\ref{fig:science}), assuming a
natural guide star, in SCAO mode.  The PSFs were calculated using the
analytic code PAOLA \citep{Jolissaint06}. This exercise has been done
in both I and K$_s$ filters to compare the impact of anisoplanatism on
photometric accuracy over the range of available wavelengths.

The main difference with SCAO images, is the presence of the strong
variation of the PSF with field position. Using a version of
Starfinder that has been adapted to SCAO images \citep{Origlia08}, we
found that the photometry becomes increasingly inaccurate and less
sensitive with increasing distance from the guide star (see
Fig.~\ref{fig:scao}).  This is largely because the light of the stars
becomes more elongated, and spread out over more pixels which reduces
the sensitivity and the spatial resolution.  The degradation in the
CMD quality and depth as we go off-axis, seen in Fig.~\ref{fig:scao},
is mainly due to increasing errors in I photometry.  The photometric
accuracies remain relatively small for K$_s$ band even 15 arcsec from
the guide star. This suggests that J-K is a better combination for
making accurate wide field CMDs with SCAO.

%Figure 16
\begin{figure}
\centering
 \includegraphics[width=9cm]{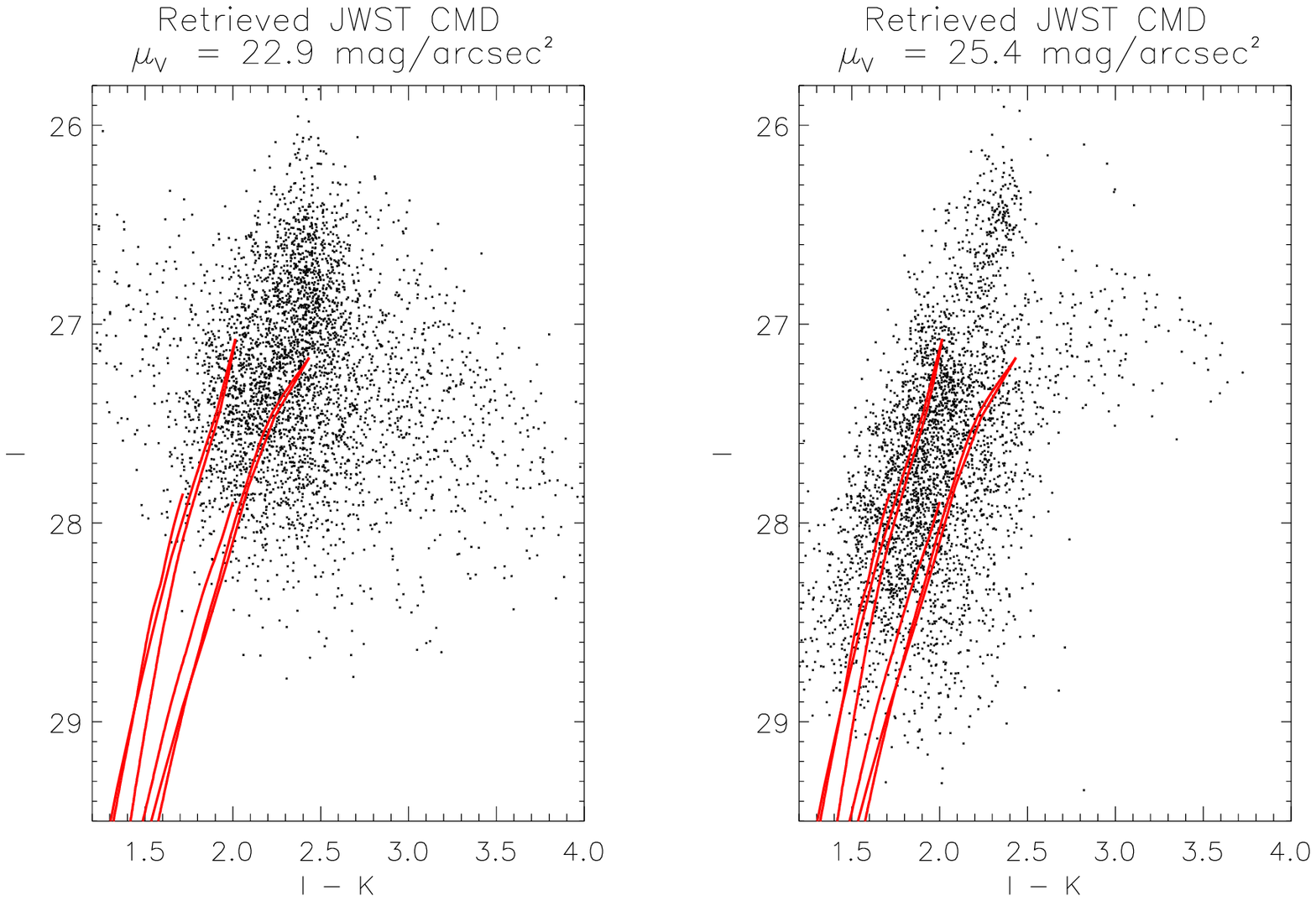}
 \caption{
These are the CMDs obtained from simulations of JWST/NIRCam images, at
two different surface brightness, $\mu_V \sim 22.9$ mag/arcsec$^2$
and 25.5 mag/arcsec$^2$.  The input CMD, and the red isochrones are
the same as in Fig.~\ref{fig:cmd_elt_k}.  The field of view simulated
is 10 times larger than for MICADO/MAORY (reflecting the larger pixel
size), and these CMDs come from 7.5~arcsec square images.
}
\label{fig:cmd_jwst}
\end{figure}

\section{Comparison with other facilities}

In the previous section we have described simulations which have
allowed us to assess the capabilities of an instrument like MICADO
coupled with an AO module like MAORY on a 42m diameter E-ELT applied
to the specific science case of the resolved stellar population of an
old galaxy at the distance of Virgo.

One way to assess how realistic these simulations may be, is to make a
broad comparison with what we get from MAD (Multi conjugate Adaptive
optics Demonstrator), an MCAO instrument which was tested on the VLT
in 2007 \& 2008. This comparison is necessarily qualitative, as MAD is
a very different MCAO system to MICADO/MAORY. MAD relies solely on
natural guide stars, and also the VLT is a much smaller telescope than
the E-ELT will be.

We also make a comparison with a future space based imager, NIRCam
(Near-IR Camera), on JWST (James Webb Space Telescope) a 6.5m IR
optimised space telescope which is scheduled for launch in late
$\sim$2015.  This is arguably the competition for ELT imaging.

\subsection{MAD on the VLT}\label{sec:mad}

The only MCAO system, which has been tested on the sky, is
MAD\footnote{see
http://www.eso.org/sci/facilities/develop/ao/sys/mad.html for
details.}, a proto-type instrument built for the VLT.  This experiment
was made to prove the concept of MCAO with natural guide stars
\citep[see][for details]{marchetti06}.  This demonstrator had three
optical Shack-Hartmann wavefront sensors for three natural guide stars
with a limiting magnitude V$\sim$13 mag. These stars were required to
be located, ideally, in the vertexes of an equilateral triangle within
a field of 2 arcminutes diameter.  The MAD engineering grade detector
had a pixel scale of 0.028 arcsec per pixel over a $\sim$1~arcmin
square field.  Several studies of Galactic stellar fields have been
published from this successful experiment
\citep[e.g.,][]{momany08,moretti09,bono09,ferraro09,sana10}.  The
Large Magellanic Cloud (LMC) was the most complex and crowded
extra-galactic stellar population studied with MAD \citep[e.g.,][and
Fiorentino et al. 2011, submitted]{campbell10}.  All these studies
showed the relative ease of use of MAD for accurate and deep IR
photometry of point sources in crowded fields. The main limitation has
been to find suitable asterisms around interesting science targets.

The difficulties in comparing directly with MICADO/MAORY lie in the
large variation in performance that result from relying on a suitable
configuration of three natural guide stars.  However, an important
technical result from MAD has been to reach the diffraction limit of
the VLT in K$_s$ band, 0.07 arcsec \citep[e.g.,][]{falomo09}. In H
band the performance has also been very good, regularly achieving
images within a factor two of the diffraction limit (0.05 arcsec).
For all the projects it was found that even if the observing
conditions were rarely optimum the correction achieved with MAD was
always an improvement over seeing-limited imaging and also over SCAO
imaging (Fiorentino et al. 2011, submitted).

It is also encouraging that standard the photometry package DAOPHOT
worked very well on MAD, and was also able to trace the PSF variation
across the field.  The MAD photometry studies (e.g., Fiorentino et
al. 2011) also concluded that the techniques applied so far to
interpret the star formation history of a galaxy, observed in V and I,
can be confidently applied to the type of data sets that we are likely
to collect with E~-ELT, i.e. IJHK$_s$ filters.  Thus a combination of
optical and IR filters is also preferred for the resolved stellar
population observations with MAD.

MICADO/MAORY will of course be a much more efficient system, first and
foremost because the laser guide stars will provide greater stability
in the AO correction, but also because many of the problems with MAD
are related to the fact that it is a test facility optimised only in
K$_s$.  MAD has proven MCAO as a viable concept that works very well,
and consistent with theoretical expectations. This gives us further 
confidence that our simulations are realistic.

\subsection{NIRCam on JWST}

%Table 4
\begin{table}
\centering
%\begin{minipage}{100mm}
\caption{Filter characteristics in Vega magnitudes for NIRCam on 
JWST}
\begin{tabular}{@{}lllll@{}}
\hline
Filters  & I & J & H & K$_s$ \\

\hline
Filter centre ($\mu$m) &	0.900	& 1.15	& 1.50	& 2.0 \\
Filter width	 ($\mu$m)	&0.225	&0.29	&0.38	&0.5\\
Background	 	&0.1 	&0.2 	&0.6	&0.1 \\
(e$^-$/s/pixel)&&&&\\

\hline
\label{table:filter_param_JWST}
\end{tabular}
%\end{minipage}

\end{table}

A potential competitor for MICADO/MAORY imaging will be
NIRCAM\footnote{http://ircamera.as.arizona.edu/nircam/} on JWST
\citep{Rieke05}. A telescope in space always has the huge advantages
of image stability and low sky background compared to terrestrial
facilities.  NIRCam is a designed to work over the 0.6 to 5 micron
wavelength range.  It will cover a larger field of view than MICADO
(2.2 arcmin square). However it will have a much larger pixel scale,
of 31.7 mas.  NIRCam is predicted to have similar sensitivity in
IJHK$_s$ filters to MICADO/MAORY. This is because the small collecting
area of JWST (6.5m diameter) is compensated by the extremely low
background flux in space (see Table\ref{table:filter_param_JWST}).
However, the JWST primary mirror size means that the diffraction limit
it considerably larger than an E-ELT, and the pixel scale reflects
this. This means that MICADO/MAORY can resolve individual stars at
significantly higher surface brightness than NIRCam.

We simulated JWST/NIRCam images for the I and K$_s$ filters in the
same way as for MICADO/MAORY, see Section~\ref{subsec:simulation},
using the same tools. We used the technical specifications found on
the JWST/NIRCam public
web-pages\footnote{http://ircamera.as.arizona.edu/nircam/features.html},
see Table~\ref{table:filter_param_JWST}.  The PSFs were made using the
JWPSF software tool \citep{Cox06}.  The NIRCam PSFs provided by this
tool are oversampled by a factor of 4, and they had to be resampled to
match pixel scale of NIRCam. We created images with the same 1~hour
integration time used for all the E-ELT simulations.  We also carried
out photometry of images using Starfinder and combined the
measurements from the I and K$_s$ filter images to make CMDs (see
Fig.~\ref{fig:cmd_jwst}).  It is very clear, that the surface
brightness regions that we can photometer are much more limited for
NIRCam, for the same stellar population at the same distance. This is
of course due to the order of magnitude difference in the diffraction
limit between the two instruments.

Comparing the MICADO/MAORY and JWST/NIRCam photometry is not so
straight forward because the highest surface brightness simulations
for which we could still carry out photometry with JWST/NIRCam images
is $\mu_V$ = 22.9 mag/arcsec$^2$, which is lower than the lowest
surface brightness considered for MICADO/MAORY (21 mag/arcsec$^2$).
This surface brightness which is uncrowded for MICADO/MAORY has severe
crowding in the JWST/NIRCam image.  At $\mu_V = 22.9$ 
mag/arcsec$^2$ in the I filter with NIRCam, we find a photometric
accuracy of $\pm$0.25~mag is reached with a star of apparent magnitude
I$=$27.1, and for MICADO/MAORY, this limit is I$=$29.3.  Thus, for the
same conditions MICADO/MAORY will be able to detect stars $>$2
magnitudes deeper in I band than JWST/NIRCam. The situation is less
dramatic for the K$_s$ filter, where for the same surface brightness,
K$=$26.2 is the limit for MICADO/MAORY and K$_s=$25.6 for JWST/NIRCam.
The difference is only $\sim$0.5 magnitudes deeper for MICADO/MAORY.
This is because of the dramatically lower background in space for K$_s$.
This whole comparison is less clear cut if uncrowded images for both
instruments are compared. That is at a surface brightness of $\mu_V$ =
25.4 mag/arcsec$^2$ for JWST/NIRCam, I=28.8 is reached with a
photometric accuracy of $\pm~0.25$mag, which is within 0.5~magnitude
of the MICADO/MAORY uncrowded limit for the same photometric error.
For the same, uncrowded surface brightness in K$_s$, the JWST/NIRCam limit
is K$_s=$27.25, which is more than $\sim$0.5~mag deeper than
MICADO/MAORY uncrowded limit.  However these uncrowded JWST/NIRCam
images will only be possible for very low surface brightness systems
(like dwarf irregular galaxies) and the distant outskirts of massive
Elliptical and Spiral galaxies at the distance of Virgo.  It should
also be kept in mind that the filters of JWST/NIRCam are also some
what different to those of MICADO/MAORY, compare
Tables~\ref{table:filter_param_JWST} and ~\ref{table:filter_param}.

The different surface brightness that can be photometered by E-ELT
(with MICADO/MAORY) and JWST (with NIRCAM) in I and K$_s$ filters is shown
in Figs.~\ref{fig:elt_jwst}.  Fig.~\ref{fig:elt_jwst} shows that only
MICADO/MAORY can carry out photometry for surface brightness, $\mu_V <
$ 23 mag/arcsec$^2$, for a reasonable photometric accuracy ($\pm$0.25)
and sensitivity. In Fig.~\ref{fig:elt_jwst} the sensitivities for both
E-ELT and JWST fall off sharply at the limit of the image crowding
mostly due to the increase in the background caused by increasing
numbers of unresolved star below the detection limit for increasing
surface brightness.

Of course for JWST/NIRCam the image is quality is likely to be much
better defined and much more stable than for MICADO/MAORY, however the
sensitivities and completeness measures of crowded JWST/NIRCam images
show that the major effect is the unresolved background flux which only
a higher spatial resolution can resolve out.

%Figure 17
\begin{figure}
\centering
 \includegraphics[width=9cm]{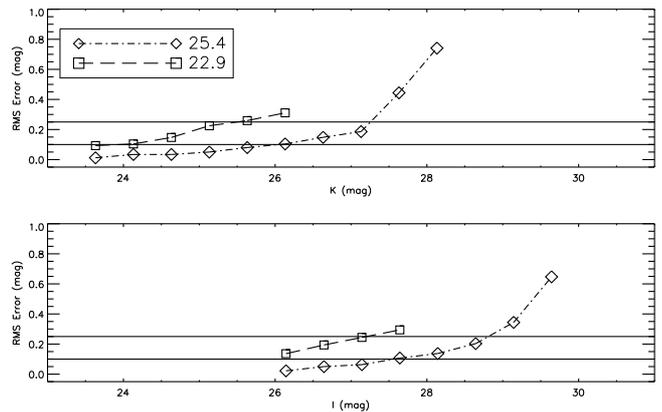}
 \caption{
Photometric errors, defined as the difference between the input and
the retrieved magnitudes, for JWST/NIRCam simulations in I and K$_s$
filters, at the maximum surface brightness where photometry can still
be achieved ($\mu_v \sim 22.9$ mag/arcsec$^2$) and also for an
uncrowded case ($\mu_v \sim 25.4$ mag/arcsec$^2$) in I and K$_s$
filters.
}
\label{fig:errors_jwst}
\end{figure}

\section{Interpretation}

\subsection{Stellar Photometry with MCAO Images}

Ancient resolved stellar populations have up to now
predominantly been studied at optical wavelengths.  This
is because the sensitivity of telescopes and instruments has always
been higher in the optical, and also old metal-poor stellar
populations tend to be relatively blue on the RGB,
compared to metal rich stellar populations of any age, and hence they
are most efficiently studied at optical wavelengths (see
Fig.~\ref{iso}).  However, there have also been specific cases
where the IR is used, for example, regions of high or variable dust
extinction.  There is presently a move towards IR optimised
facilities. This is partly driven by the demands of detecting and
studying extremely high redshift galaxies, but it is also where it is
possible to correct for atmospheric effects on ground-based image
quality.  Both JWST and E-ELT are optimised for the IR. For E-ELT this
is determined by the technology limitations of being able to make
sufficiently accurate AO corrections.

Here we have presented a very specific technical study of a single
resolved stellar population, to assess the feasibility of obtaining
accurate photometry with MICADO/MAORY (in I, J, H, K$_s$ filters) for
resolved stars in crowded fields at the distance of Virgo.  The main
advantage that the MICADO/MAORY instrument will have over all others
being planned at present, including NIRCam on JWST, is the extremely
high spatial resolution it should be capable of at optical/IR
wavelengths.  This provides a unique opportunity to look at resolved
stellar populations deep in the heart of Elliptical 
galaxies (the best examples of which are in the Virgo cluster, at
17~Mpc distance).  Thus, the primary aim of this study was to see how
well stars could be resolved and photometered in much higher surface
brightness (crowded) conditions than are feasible at present.  The
conclusion is that this is a rather challenging, but viable, aim.

%Figure 18
\begin{figure}
\resizebox{\hsize}{!}{\includegraphics{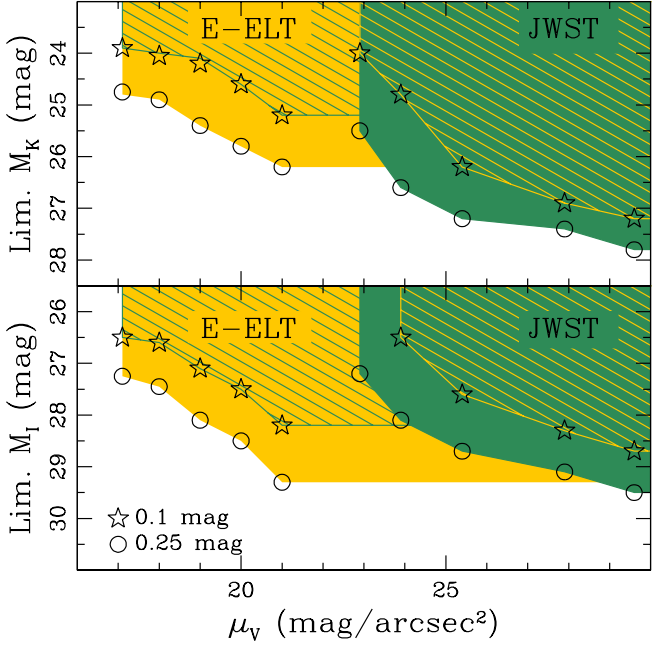}}
 \caption{
Here the surface brightness and limiting magnitude limits are shown
for E-ELT (MICADO/MAORY) photometry and JWST (NIRCam) photometry of
resolved stars in an old galaxy at the distance of Virgo (17~Mpc).
The open star symbols demarcate the sensitivity for photometric errors
$\pm$0.1~mag and the open circles for more conservative photometric
errors of $\pm$0.25~mag.
}
 \label{fig:elt_jwst}
\end{figure}

We have shown that despite the very peculiar PSF (see
Fig.~\ref{fig:PSFs}), especially in the I filter, accurate
photometry can be carried out even with current standard photometry
packages. However, there is certainly room for improvement to optimise
photometry algorithms to deal with complex MCAO PSFs, as the
photometric errors are still relatively large, even at bright
magnitudes.  It will also undoubtedly be possible in the future to
push the accuracy of photometry in crowded regions using PSF
reconstruction techniques. The better the PSF properties are
known the more precisely it is possible to remove bright stars from
an image and more accurately photometer the fainter stars under their
wings.

We have found that it is possible to reach much fainter magnitude
limits in I, in the same exposure time, than in J, H or K$_s$ and the
deepest luminosity functions will be obtained in this filter. This is
a slightly surprising result given that the AO correction is much less
effective in I, compared to J and K$_s$.  This means that not even the
much better AO performance, and better defined and more sharply peaked
PSFs in the IR filters can compensate for the difference in sky
brightness that drives the basic sensitivity difference.  For
example in the J filter the strehl is a factor $\sim$3 better than in
the I filter (see Table~\ref{table:ee-sr}).  This means that the
signal in the central aperture of the PSF is $\sqrt{3}$ better than in
I. However, the background in J is 16.5~mag; and in I it is 19.7~mag,
which is a difference of $\sim$3~mag. These effects do not balance
out, and so the sky background dominates, with the result that the I
filter is significantly more sensitive than J, H or K$_s$.

It has long been known that the larger is the colour baseline, the
more spread out is the RGB (see Fig.~\ref{iso}). In principle this
should make it easier to interpret the CMD more accurately in terms in
age and metallicity using I-K colours instead of J-K.  As
Figs.~\ref{lim-cmds} and \ref{errors-k} show, this is not such a
clear-cut argument as the sensitivity plots might
suggest. Fig.~\ref{lim-cmds} shows that strictly speaking I-K is more
detailed than J-K, but not with a huge significance.
Fig.~\ref{errors-k} shows how clearly the choice of the best
filter combinations depends upon the colours of the stars being
observed. The redder the stars the larger has to be the difference
between the magnitude limit in I and K$_s$ for the sensitivities to be
matched. In fact the average I-K colour of the RGB in our simulations,
which is probably quite blue for a typical giant Elliptical
galaxy, ranges between 2 and 3, which matches quite well the
sensitivity difference between I and K$_s$ for MICADO/MAORY. If the
K$_s$ sensitivity were to increase this would not affect {\it these}
observations. In the case of a much redder stellar population
(I-K$>$2.8), then the strongest limitation on the depth of the CMDs
would be the I magnitude (see Fig.~\ref{errors-k}).  This suggests
that a good observing strategy for studying Elliptical galaxies is to
use three filters, namely I, J \& K$_s$. The I-K combination will be
most useful for studying the metal poor population, or to determine if
one is present. The J-K colour will be more revealing about the
presence of extremely red evolved stars, such as Carbon stars and
metal rich AGB stars.

An important question to ask, based on these simulations: Is the I
filter crucial for the science case addressed in this paper?  Does it
lead to significant enough improvement to scientific return to warrant
the large amount of effort to install it?  The answer is not a
straight forward yes or not. It is certainly the most sensitive filter
for the observations of the bulk of the resolved stellar populations,
and especially for metal poor stars.  However, given the
fundamental limits on JHK sensitivity for the specific case of
photometering the RGB of old stellar populations in galaxies in
Virgo, there is no resason to prefer one filter over the other. However, for CMD analysis we need a colour for the most accurate
analysis.  Thus at present, for this science case, the answer is that
care should be taken that the I filter PSF and sensitivity is
delivered as has been presented now. The PSF from MAORY has been
provided on a ``best effort'' basis, and it is at the limits of what
is thought to be possible given the current technical specifications
of both MAORY and the telescope adaptive mirror (M4).  The I-K colour
is more spread out than J-K, and this may often make it easier to
disentangle complex stellar populations more accurately in I-K, see
Fig.~\ref{lim-cmds} in certain cases. In conclusion the I filter
should remain part of the ``standard'' filter set of any MCAO system
for the E-ELT, even with a minimal AO performance.

A possible upgrade path for the telescope may allow better AO
performance in the optical which could lead to deep I luminosity
functions, which can also be used to interpret the properties of
distant stellar populations. If the sensitivity of the I filter can be
pushed to reach the Horizontal Branch limit in Virgo (I$\sim$31), then
it will be possible to directly and unambiguously determine the
presence of an ancient and/or metal-poor stellar population in
Elliptical galaxies in the Virgo cluster.

\subsection{What we may learn about Elliptical  galaxies}

\begin{figure}
\resizebox{\hsize}{!}{\includegraphics{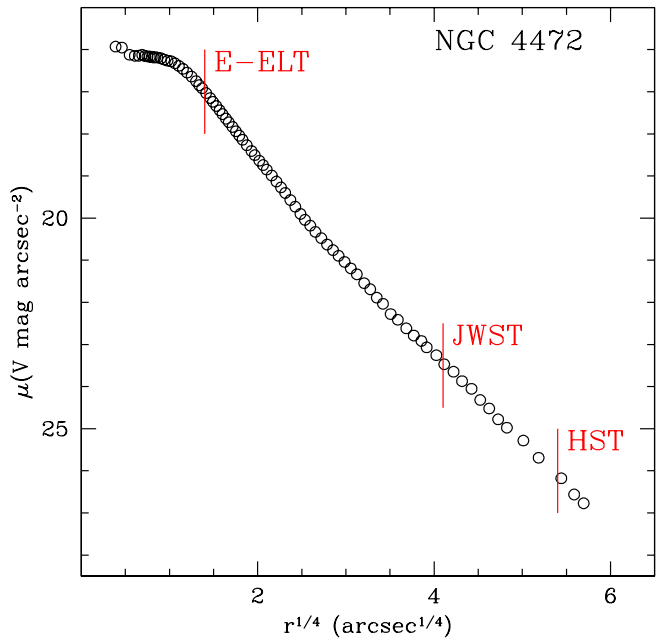}}
 \caption{
Here we plot the surface brightness profile of NGC~4472
\citep[from][]{Kormendy09}, a ``typical'' giant Elliptical galaxy in
the Virgo cluster.  There is a vertical line placed at the distance
from the centre of the galaxy at the highest surface brightness for
which CMDs can be made and photometered with E-ELT and JWST (coming
from this study), and HST/ACS \citep{Durrell07}.
}
 \label{fig:science2}
\end{figure}

The centre of a giant Elliptical galaxy in Virgo typically has a
central surface brightness, $\mu_v \sim 16$~mag/arcsec$^2$.  Our
simulations suggest MICADO/MAORY will be able to detect individual
stars, even old, metal poor stars, close to the central regions
of such a galaxy.  It will be able to make CMDs within 5~arcsec of the
centre, of even the largest Elliptical galaxies, compared to
250~arcsec for JWST, as is shown in Fig.~\ref{fig:science2}.

%Figure 19

Of course, as can be seen from Fig.~\ref{fig:cmd_elt_k} the amount of
information that can be extracted from a CMD at a surface brightness,
$\mu_v = 17$~mag/arcsec$^2$ is likely to minimal. Assuming that the
distance to the galaxy is very well known, than it will be possible to
tell if we are looking at the AGB star population.  If the distance is
uncertain, and it is not known a priori if a stellar population
contains AGB stars or not, then there will be no clear way of deciding
if AGB or RGB stars are being detected from this kind of CMD.

All the CMDs in Fig.~\ref{fig:cmd_elt_k} are still only of the upper
region of the RGB in a Virgo galaxy, so it is not always going to be
possible to disentangle the properties of complex
populations. However, in the case of distinct age or metallicities
differences such as we used for our simulations, it will be possible
to infer the presence of two populations with different
characteristics.  It should also be possible to determine the age and
metallicity spread (see Fig.~\ref{iso}). But these are not the ideal
CMDs to make a unique interpretation \citep[e.g.,][]{Gallart05}.  For
the kind of relative study that will be possible well populated CMDs
are crucial, and thus a large field view (compared to the size of the
galaxy being studied) is very important. The major advantage of
an MCAO imager for this science case, is the wide
field of view that is possible ($\sim$53 arcsec square), with 
uniform AO correction. This allows an accurate and detailed comparative
study of a significant area of most Elliptical galaxies at the
distance of Virgo.

It is almost certainly useful to consider the constraints that may
come from comparing low surface brightness regions (where the
photometry is more accurate) to higher surface brightness regions in
the same galaxies. This can be done using detailed modelling to be as
accurate as possible in comparing CMDs to make sure that the effects
of crowding are well understood. It should not be forgotten that the
CMDs presented, in Fig.~\ref{fig:cmd_elt_k} are for extremely small
fields of view (0.75~arcsec squared). When the full field of MICADO is
considered 5000 times more stars will be included in each CMD. These
are likely to cover a range of surface brightness going from the inner
to the outer regions of a typical Elliptical galaxy at the distance of
Virgo. This will allow comparative studies of how the numbers of stars
of different colours and luminosity vary with position, and 
to study the relative importance of different
types of stars on the RGB at different positions in a range of
Elliptical galaxies. This will provide valuable information as to how
uniform the stellar population is for a given system. The surface
brightness distribution is typically very smooth for Elliptical galaxies, 
but for theses types
of stars, a global colour and luminosity can hide a lot of variation
\citep[e.g.,][]{Monachesi11}. Comparing the resolved stellar
populations of the inner and outer regions of Elliptical galaxies will
have important implications for formation scenarios. It will
also be possible to study the numerous population of lower surface
brightness Ellipticals in Virgo in their entirety.

\begin{figure}
\resizebox{\hsize}{!}{\includegraphics{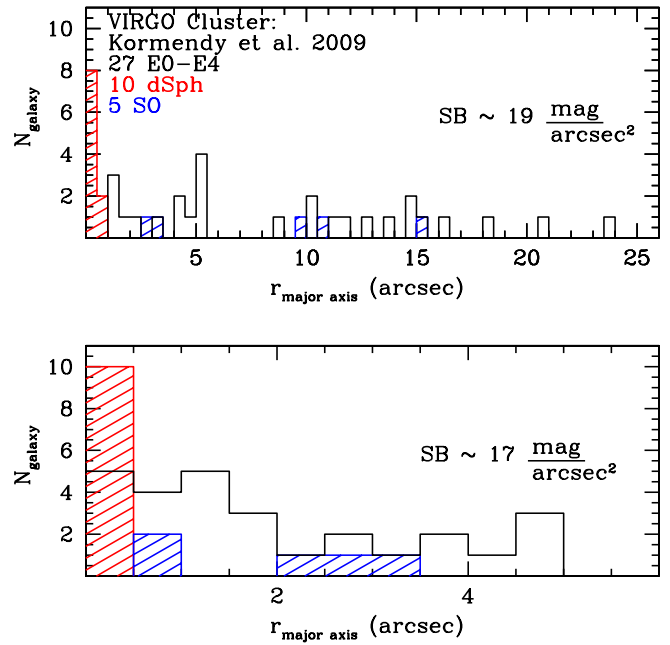}}
 \caption{
For a sample of galaxies in the Virgo cluster \citep{Kormendy09} we
plot the position on the major axis (in arcsec) to which they can be
resolved into individual stars using MICADO/MAORY at two different
surface brightness limits (SB), given in each plot. The
same sample of galaxies is plotted in each case; 10 dwarf spheroidals
(in red); 27 Elliptical galaxies (black) and 5 S0 galaxies (blue).
}
 \label{fig2:science}
\end{figure}

Fig.~\ref{fig2:science} shows, from the surface brightness limits of
MICADO/MAORY imaging, how many galaxies in Virgo
\citep[from][]{Kormendy09} can be studied in detail, to which distance
from the centre, depending on the surface brightness limit chosen.
Fig.~\ref{fig2:science} shows that it will be possible to resolve
the entire stellar population of dwarf spheroidal galaxies, as they
are uniformly low surface brightness systems.  It should be realised
that in most of these cases the galaxies are no more than 1$-$3~arcmin
across, at a surface brightness of $\sim$22mag/arcsec$^2$.  The wide
field of MICADO/MAORY will thus efficiently allow the sampling of a range of
surface brightness across each galaxy.

Of course MICADO/MAORY will be able to make a more detailed CMD
of closer by giant Elliptical galaxies, such as NGC~3379 (M~105) the
dominant member of the Leo group, at a distance of 10.5~Mpc
\citep{salaris98}, which corresponds to a distance modulus of,
(m-M)$_0 = 30.3$. This is thus 0.7~magnitude closer than the
simulations at the distance of Virgo 
shown here. This means that the CMDs will go 0.7
magnitudes deeper in each filter for the same photometric errors, at
the same surface brightness. The central surface brightness of
NGC~3379 is about $\mu_V = 19$ mag/arcsec$^2$. So we can look in the
centre of this galaxy and go 0.7 mag deeper than the 19 mag/arcsec$^2$
(and fainter surface brightness) CMDs shown in Fig.~\ref{fig:cmd_elt_k} 
in one hour of integration time. 

Fig.~\ref{errors-k} allows the reader to estimate the impact of
different sensitivity limits and stellar colours on the final
photometric errors in both magnitude and colour.

%Figure 20

\section{Conclusions}

In this paper we have carefully simulated MICADO/MAORY images of an
old resolved stellar population at the distance of Virgo
in four broad band filters (I, J, H \& K$_s$).  These simulations are
based upon the most detailed information about the telescope and
instrument characteristics currently available.  We restricted
ourselves to one narrowly defined science case, for one stellar
population. Keeping this aspect of the study fixed allowed us to
restrict the number of variables that had to be considered, hopefully
leading to a clearer understanding of the the importance of
observational and instrumental effects, such as sky brightness,
seeing, strehl and PSF shape.  There is certainly scope for expanding
this study to other types of stellar populations, especially those at
closer distances where it will be possible to make accurate CMDs of
the Main Sequence Turnoff region.  For our science case we found that
reaching the tip of the RGB at the distance of Virgo is clearly
feasible in two filters
in observations of one hour per filter, when the surface brightness is
no higher than 17~mag/arcsec$^2$.  Deeper CMDs, in the same exposure
time, are possible for a surface brightness of less than
20~mag/arcsec$^2$, where CMDs can reach $\sim$2~mags below the tip of
the RGB.  Thus, our simulations suggest that obtaining accurate CMDs
for resolved stellar populations in Elliptical galaxies at the
distance of the Virgo cluster is a challenging goal, but feasible.

We suggest that the best strategy for resolving and photometering
individual old stars at the distance of Virgo is to
use three filters, namely I, J \& K$_s$. The I-K combination will be
most useful for studying the metal poor population, or to determine if
one is present. The J-K colour will be more revealing about the
presence of extremely red evolved stars, such as Carbon stars and
metal rich AGB stars.

We also performed simulations in a similar fashion for the NIRCam
imager on JWST. This will clearly work at a different (barely
overlapping) surface brightness range to MICADO/MAORY.  Of course the
wider field of view of NIRCam is better suited than MICADO/MAORY for
studying the properties of the extended halo of large galaxies.
From this comparison we can infer that MICADO/MAORY and NIRCam will be
a useful complementary pair for detailed studies of giant 
Elliptical-like 
galaxies in Virgo.  MICADO/MAORY will be able to probe the high
density central parts, and NIRCam, with a larger field of view but
fainter surface brightness limit, will be more effective observing the
extensive outer regions.  Thus the two instruments will allow us to
study large and small scale properties of the resolved stellar
populations in a large number of Elliptical galaxies of all sizes and
characteristics in Virgo.

Thus, our simulations have led to deeper insights into the AO
performance issues of an MCAO images on an E-ELT, and can be used to
make a useful instrument for studies of stellar populations and also
to improve the software to deal with MCAO images.

\section*{Acknowledgments}

This paper grew out of discussions during the E-ELT MICADO Phase A
study, and was expanded after the completion of the MAORY Phase A
study.  We thank Enrico Marchetti for useful discussions and comments
on an early version of this manuscript. We thank Renato Falomo for
interesting discussions during the MICADO Phase A study.  We are
grateful for funding for this project from an NWO-VICI grant (AD, GF
\& ET), as well as ESO funding for MICADO \& MAORY phase A studies.

% for the bibliography, at the end
\bibliographystyle{aa} % style aa.bst
\bibliography{et-ao2} % your references Yourfile.bib

\end{document}